\documentclass[prl,twocolumn, showpacs, groupedaddress,nopacs]{revtex4}  

\usepackage{graphicx}
\usepackage{dcolumn}
\usepackage{bm}
\usepackage[bbgreekl]{mathbbol}
\usepackage{amsmath}
\usepackage{amsfonts}
\usepackage{xcolor}
\usepackage[mathscr]{euscript}

\def\ii{{\rm i}} 
\def\rb{\textbf{r}} 
\def\db{\textbf{d}} 
\def\ge{\sigma_{ge}}  
\def\eg{\sigma_{eg}}  
\def\braket#1{\mathinner{\langle{#1}\rangle}}

\begin{document}
\title{Atom-atom interactions around the band edge of a photonic crystal waveguide}
\author{J. D. Hood$^{1,2}$, A. Goban$^{1,2,\dag}$, A. Asenjo-Garcia$^{1,2}$, M. Lu$^{1,2}$, S.-P. Yu$^{1,2}$, D. E. Chang$^{3}$, and H. J. Kimble$^{1,2,\ast}$
\footnotetext{\small $^{\dag}$ Present address: JILA, University of Colorado, 440 UCB, Boulder, Colorado 80309, USA} 
\footnotetext{\small $^{\ast}$ hjkimble@caltech.edu}}

\address{$^1$ Norman Bridge Laboratory of Physics MC12-33}
\address{$^2$ Institute for Quantum Information and Matter, California Institute of Technology, Pasadena, CA 91125, USA}
\address{$^3$ ICFO-Institut de Ciencies Fotoniques, The Barcelona Institute of Science and Technology, 08860 Castelldefels (Barcelona), Spain}

\date{\today}
\begin{abstract}
Tailoring the interactions between quantum emitters and single photons constitutes one of the cornerstones of quantum optics. Coupling a quantum emitter to the band edge of a photonic crystal waveguide (PCW) provides a unique platform for tuning these interactions. In particular, the crossover from propagating fields $E(x) \propto e^{\pm ik_x x}$ outside the bandgap to localized fields $E(x) \propto e^{-\kappa_x |x|}$ within the bandgap should be accompanied by a transition from largely dissipative atom-atom interactions to a regime where dispersive atom-atom interactions are dominant. Here, we experimentally observe this transition for the first time by shifting the band edge frequency of the PCW relative to the $\rm D_1$ line of atomic cesium for $\bar{N}=3.0\pm 0.5$ atoms trapped along the PCW. Our results are the initial demonstration of this new paradigm for coherent atom-atom interactions with low dissipation into the guided mode.
\end{abstract}
\maketitle

Recent years have witnessed a spark of interest in combining atoms and other quantum emitters with photonic nanostructures \cite{LMS15}. Many efforts have focused on enhancing emission into preferred electromagnetic modes relative to vacuum emission, thereby establishing efficient quantum matter-light interfaces and enabling diverse protocols in quantum information processing \cite{K08}. Photonic structures developed for this purpose include high-quality cavities \cite{VY03,YSH04, ADW06, HBW07, TTL13}, dielectric fibers \cite{BHB09,BHK04,LVB09,VRS10,GCA12,polzik}, metallic waveguides \cite{CSD07,AMY07,HKS11}, and superconducting circuits \cite{WSB04,VFL13,DS13}.  Photonic crystal waveguides (PCWs) are of particular interest since the periodicity of the dielectric structure drastically modifies the field propagation, yielding a set of Bloch bands for the guided modes \cite{JMW95}. For example, recent experiments have demonstrated superradiant atomic emission due to a reduction in group velocity for an atomic frequency near a band edge of a PCW \cite{GHH15}.

A quite different paradigm for atom-light interactions in photonic crystals was proposed in Refs.~\cite{Y1987,J1987,JW1990, K1990}, but has yet to be experimentally explored. In particular, when an atomic transition frequency is situated within a bandgap of a PCW, an atom can no longer emit propagating waves into guided modes (GMs) of the structure. However, an evanescent wave surrounding the atoms can still form, resulting in the formation of  atom-photon bound states \cite{shi,calajo}. This phenomenon has attracted new interest recently as a means to realize dispersive interactions between atoms without dissipative decay into GMs. The spatial range of atom-atom interactions is tunable for $1$D and $2$D PCWs and set by the size of the photonic component of the bound state \cite{DHH15,GHC15}. Many-body physics with large spin exchange energies and low dissipation can thereby be realized, in a generalization of cavity QED arrays  \cite{HBP06, GTC06}. Fueled by such perspectives, there have been recent experimental observations with atoms \cite{GHH15,GHY14,YHM14} and quantum dots \cite{LSJ08,YTB15} interacting through the GMs of photonic crystal waveguides, albeit in frequency regions outside the bandgap, where GMs are propagating fields.

\begin{figure*}
\begin{center}
\includegraphics[width=\linewidth]{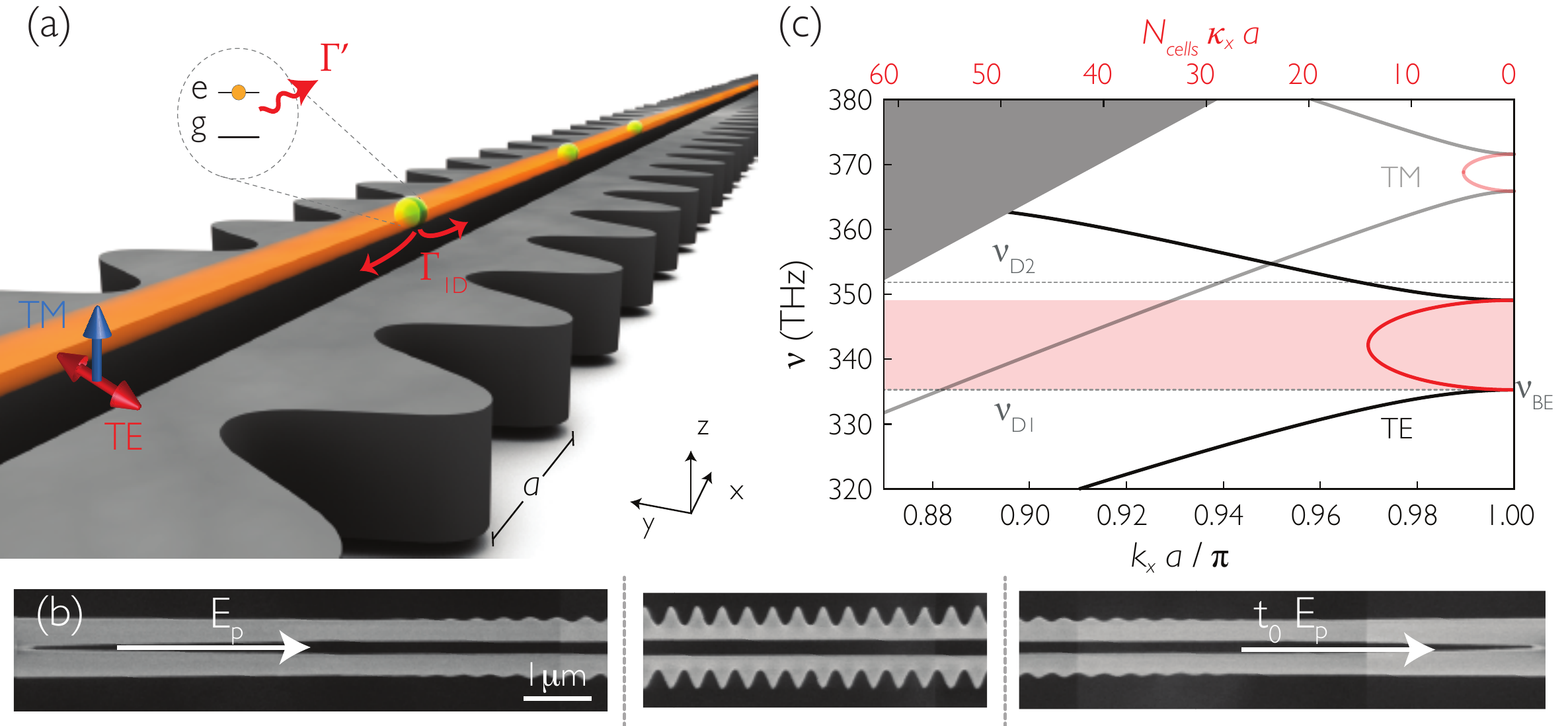}
\caption{Description of the alligator photonic crystal waveguide (PCW). {\bf (a)} 
Atoms are trapped above the PCW in an optical dipole trap formed by the reflection of a near normal-incidence external beam \cite{GHH15}. The orange cylinder represents the confinement of the atoms, which is $\Delta x_{\rm A} \simeq \pm 6$~$\mu$m along the axis of the device, and $\Delta y_{\rm A} \simeq \Delta z_{\rm A}  \simeq \pm 30$~nm in the transverse directions~\cite{SM}.
 The three green spheres represent trapped atoms that interact radiatively via the fundamental TE guided mode, polarized mainly along $y$. The decay rate for a single atom into the PCW is $\Gamma_{\text{1D}}$ (red arrows), and the decay rate into all other modes is $\Gamma'$ (wavy red). {\bf (b)} SEM images of portions of the tapering and PCW sections. The suspended silicon nitride device (grey) consists of $150$ cells and $30$ tapering cells on each side. The lattice constant is $a=370$~nm and thickness is 185~nm. {\bf (c)} Calculated band structure of the fundamental TE (solid) and TM (translucent) modes using an eigenmode solver \cite{COMSOL} and the measured SEM dimensions, which are modified within their uncertainty to match the measured bands. The black curves represent the Bloch wave-vector $k_x$ (lower axis).  The red curves show the attenuation coefficient $\kappa_x$ of the field for frequencies in the bandgap (upper axis), and are calculated by means of an analytical model \cite{SM}. The dotted lines mark the frequencies  of the Cs $\rm D_1$ ($\nu_{\text{D1}}=335.1$~THz) and D$_2$ ($\nu_{\rm D2}=351.7$~THz)  transitions. The dielectric band edge is indicated as $\nu_{\text{BE}}$. The pink (gray) shaded area represents the TE bandgap (the light cone).}
\label{Fig1}
\vspace{-8mm}
\end{center}
\end{figure*}

In this manuscript, we report the first observation of collective dispersive shifts of the atomic resonance  around the band edge of a photonic crystal. 
 Thermal tuning allows us to control the offset of the band edge frequency ($\nu_{\text{BE}}$) of the PCW relative to frequency $\nu_{\text{D1}}$ of the $\rm D_1$ line of cesium. In both the dispersive domain (i.e., $\nu_{\text{D1}}$ outside the bandgap with electric field $E(x) \propto e^{\pm ik_x x}$) and reactive regime (i.e., $\nu_{\text{D1}}$ inside the bandgap with $E(x) \propto e^{-\kappa_x |x|}$), we record transmission spectra for atoms trapped along the PCW, as illustrated in Fig.~1(a).

To connect the features of the measured transmission spectra to underlying atom-atom radiative interactions, we have developed a formalism based on the electromagnetic Green's function.  The model allows us to infer the peak single-atom frequency shift of the atomic resonance $J_{\text{1D}}(\Delta_{\rm BE})$ and guided mode decay rate $\Gamma_{\text{1D}}(\Delta_{\rm BE})$ as functions of detuning  $\Delta_{\rm BE} = \nu_{\text{D1}} - \nu_{\text{BE}}$ between the atomic $\nu_{\text{D1}}$ and band edge $\nu_{\text{BE}}$ frequencies.
From the observation of superradiant emission outside the bandgap, we infer the average number of trapped atoms to be $\bar{N}=3.0\pm0.5$, as described in Ref.~\cite{GHH15} and the supporting material \cite{SM}.  For frequencies inside the bandgap ($\Delta_{\rm BE}=50$~GHz) the  ratio of dissipative to coherent rates is $\mathcal{R}=\Gamma_{\text{1D}}/J_{\text{1D}}=0.05\pm 0.17$, due to the exponential localization of the atomic radiation in the bandgap.  For comparison, the prediction for our system from cavity quantum electrodynamics (CQED) models alone is $\mathcal{R_{\rm{CQED}}}=0.30\pm0.04$. Besides yielding a more favorable ratio between coherent and dissipative guided mode rates, PCWs offer significant advantages when compared to conventional cavities as platforms for atom-light interfaces. First, the range of interaction in a PCW is tunable, ranging from effectively infinite to nearest neighbor, in contrast to the fixed infinite range of a cavity. Second, due to the multimode nature of PCWs, one can employ different guided modes as different interaction channels to which the atoms simultaneously couple.

\textbf{Alligator Photonic Crystal Waveguide -} Figure~1(a) provides an overview of our experiment with atoms trapped near and strongly interacting with the TE-like mode of an alligator PCW. The suspended silicon nitride structure consists of $N_{\rm cells}=150$ nominally identical unit cells of lattice constant $a=370$~nm, and is terminated by $30$ tapering cells on each side, as shown in the SEM images in Fig.~1(b). The tapers mode-match the fields of the PCW to the fields of uncorrugated nanobeams for efficient input and output coupling. Design, fabrication, and characterization  details are described in Refs.~\cite{YHM14,GHY14,GHH15}. Figure~1(c) shows the nominal cell dispersion relations for the TE (polarized mainly along $y$) and TM-like modes (polarized mainly along $z$).  After release of the SiN structure from the Si substrate, a low power CF4 etch is used to align the lower/`dielectric' TE band edge ($\nu_{\text{BE}}$) to the Cs $\rm D_1$ transition ($\nu_{\text{D1}}$). The TM mode has band edges far detuned from the both the Cs $\rm D_1$ and $\rm D_2$ lines. In our experiment, the TE mode is used to probe the atoms, while the TM mode with approximately linear dispersion serves to calibrate the density and trap properties. 

\begin{figure*}
\begin{center}
\includegraphics[width=\linewidth]{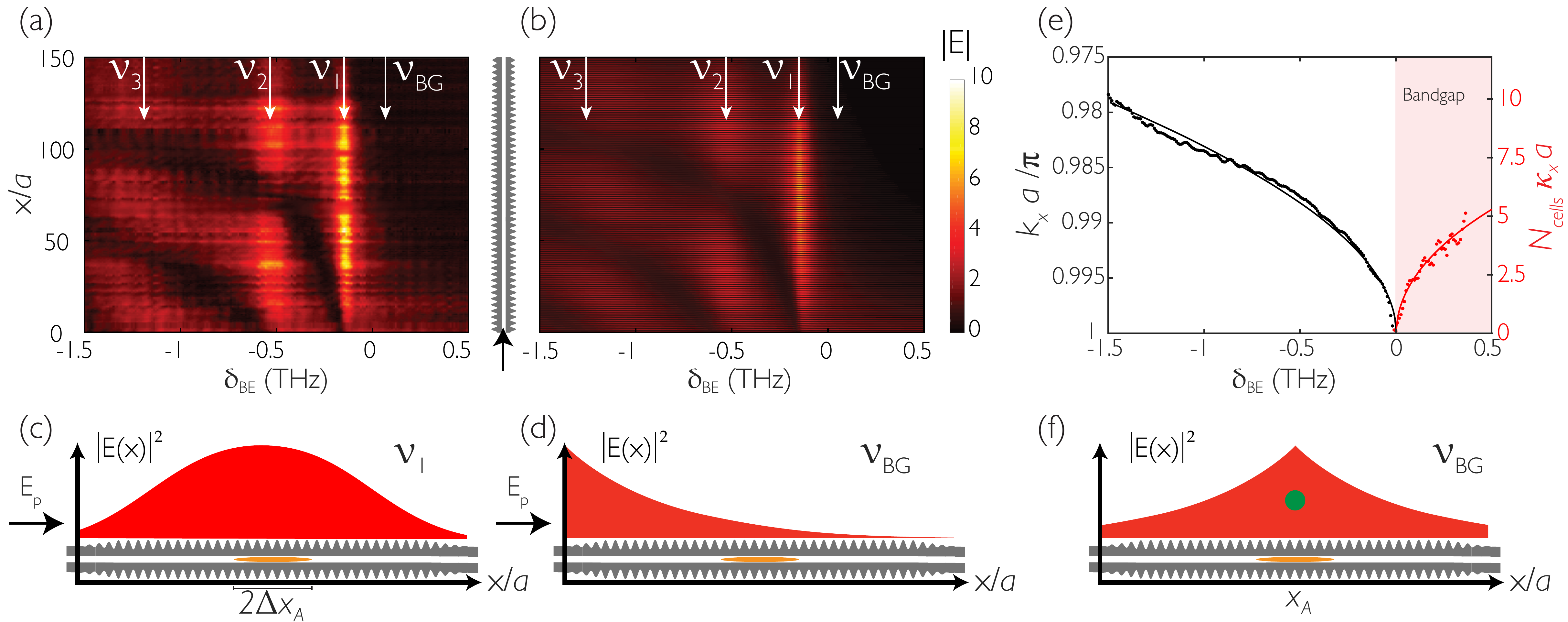}
\caption{Characterization of the alligator PCW. {\bf (a)} Measured and {\bf (b)} calculated  electric field magnitude along the PCW, as functions of position $x$ along the PCW and probe detuning $\delta_{\rm BE}=\nu_{\rm p} - \nu_{\text{BE}}$ relative to $\nu_{\text{BE}}$ for the dielectric band edge. {\bf (c,d)} Guided mode intensity $|E(x)|^2$ along PCW at two different frequencies: (c) $\nu_1$ for the first cavity resonance showing a resonant `super mode' and (d) $\nu_{\text{BG}}$ inside the bandgap displaying exponential decay ($N_{\rm cells}\kappa_x a=2.0$ at $\nu_{\rm BG}$).  For clarity, the number of cells of the nominal and tapering sections is decreased by a factor of 5, and the Bloch periodicity ($a=370$~nm), while present, is not shown in the intensity. The orange ovals represent the confinement of the atoms in the optical trap above the PCW, which is $\Delta x_{\rm A} \simeq \pm 6 $~$\mu$m along the $x$-axis of the device and $\Delta y_{\rm A} \simeq \pm 30$ nm, with a PCW gap width of $220$~nm.
 {\bf (e)} Dispersion relation for the projected wave vector $k_x$ and attenuation constant $\kappa_x$ versus probe detuning $\delta_{\rm BE}$ deduced for the PCW obtained by fitting the data in (a) to a model of the device \cite{SM}. The shaded pink area represents frequencies inside the bandgap.   {\bf (f)} Plot of the exponentially localized emission $e^{-2\kappa_x |x-x_{\rm A}|}$ from an atom (green sphere) at position $x_{\rm A}$ with transition frequency $\nu_{\text{D1}} = \nu_{\text{BG}}$ inside the bandgap.}
\vspace{-7mm}
\label{Fig2}
\end{center}
\end{figure*}

In order to better understand atomic interactions with the PCW, it is helpful to visualize the spatial profile of the fields generated absent atoms, when light is input from one end. Figure~2(a) shows the measured intensity along the length of the PCW as a function of probe detuning $\delta_{\rm BE} =  \nu_{\rm p} - \nu_{\text{BE}}$ around the band edge, where $\nu_{\rm p}$ is the probe frequency. The intensity was measured by imaging weak scatterers along the length of the alligator PCW that, after calibration, serve as local probes of the intensity \cite{SM}. Figure~2(b) shows the corresponding finite-difference time-domain (FDTD) simulated intensity \cite{Lumerical}. In both images, resonances appear at $\nu_{\rm p} = \nu_{1,2,3}$ due to the weak cavity formed by the reflections of the tapers.  The spatial modulation of the intensity at the resonances due to the cavity effect is approximated by $|E(x)|^2 \approx \cos^2(\delta k_x \, x)$, where $\delta k_x = \pi/a - k_x $  is the effective wave-vector near the band edge.  The $n$'th resonance  at frequency $\nu_n$ is such that $\delta k_x = n \pi/L$, where $L$ is the effective length of the PCW (including field penetration into the tapers).  Fig. 2(c) shows a plot of $|E(x)|^2$ for a probe input at frequency $\nu_{\rm p} = \nu_1$ at the first resonance.  Inside the bandgap ($\Delta_{\rm BE} >0$) the field is evanescent, and $\delta k_x = i \kappa_x$.  Fig. 2(d) plots $|E(x)|^2$ for probe frequency $\nu_{\rm p} = \nu_{\text{BG}}$ inside the bandgap, and shows the exponential decay of the intensity. Using a model for the field in a finite photonic crystal \cite{SM}, we fit the measured intensity for each frequency in Fig. 2(a) and Fig. 2(b) and extract $\delta k_x$ and $\kappa_x$, thereby obtaining the dispersion relations shown in Fig.~2(e). Importantly, we determine the band edge frequency for the actual device to be $\nu_{\text{BE}} - \nu_1 = 133 \pm 9$~GHz relative to the readily measured first resonance at $\nu_1$, which is in good agreement with the FDTD simulated result of $135$~GHz.

 Both $\nu_{1}, \nu_{\text{BG}}$ are relevant to our measurements of transmission spectra with trapped atoms. The presence of a `cavity' mode at $\nu_1$ implies that the emission of an atom with transition frequency $\nu_{\text{D1}} = \nu_{1}$ will generate a field inside the PCW with a similar spatial profile to that of the cavity mode, as shown in Fig.~2(c). By contrast, atomic emission in the regime with  $\nu_{\text{D1}} = \nu_{\text{BG}}$ within the bandgap will excite an exponentially localized mode centered around $x_{\rm A}$, as illustrated in Fig. 2(f).

\textbf{Experiment.}--  Cs atoms are trapped above the surface of the alligator PCW, as shown in Fig.~1(a), using a similar experimental setup to that reported in Ref.~\cite{GHH15}. As described in more detail in the previous reference, the decay rate into the guided mode $\Gamma_{\rm 1D}$ is exponentially sensitive to the trap position above the surface of the alligator PCW. Our calculations and measurements of $\Gamma_{\rm 1D}$ agree with COMSOL simulations \cite{COMSOL} of the trap position, and thus we are able to determine that the Cs atoms are trapped $125\pm 15$~nm above the surface of the alligator PCW. Atoms are cooled and trapped in a MOT around the PCW, and then loaded into a dipole trap formed by the reflection from the device of a frequency red-detuned side illumination (SI) beam. The SI beam has a waist of $50$ $\mu$m, and the polarization is aligned along the $x$ axis for  maximum reflection from the PCW.  
 We measure a $1/e$ trap life time of $\sim\! 30$ ms, and we estimate an atom temperature of $\!\sim\! 30$~$\mu$K from time-of-flight measurements.  From the trap simulations (for details see supporting materials \cite{SM}), we infer that the atoms are confined to a region $125$~nm above the surface with dimensions $\Delta x_{\rm A} \simeq \pm 6$~$\mu$m, $\Delta y_{\rm A} \simeq\Delta z_{\rm A} \simeq \pm 30$~nm. The simulations predict that more energetic atoms escape the trap and collide into the structure, since the weakest direction of the trap is along the diagonals of the $y$-$z$ plane due to Casimir-Polder forces. 

In order to estimate the average number of trapped atoms, we measure the superradiant atomic decay rate when the atom frequency $\nu_{\text{D1}}$ is tuned to the first  resonance $\nu_1$ of the PCW (Fig.~2(c)) \cite{GHH15}.  Due to the strong dissipative interactions between the atoms and with $J_{\rm 1D} \approx 0$, the collective decay rate is enhanced as compared to the single atom decay rate, and we infer an average atom number of $\bar{N} = 3.0 \pm 0.5$ \cite{SM}. 
In the low density limit $\bar{N} \ll 1$, the measured decay rate corresponds to that of a single atom.  We then measure a guided mode decay rate $\Gamma_{\rm 1D}  =(1.5\pm 0.2)\, \Gamma_0 $, which is in good agreement with the FDTD simulations at the calculated trap location \cite{SM}.

\begin{figure}[b!]
\includegraphics[width=\columnwidth]{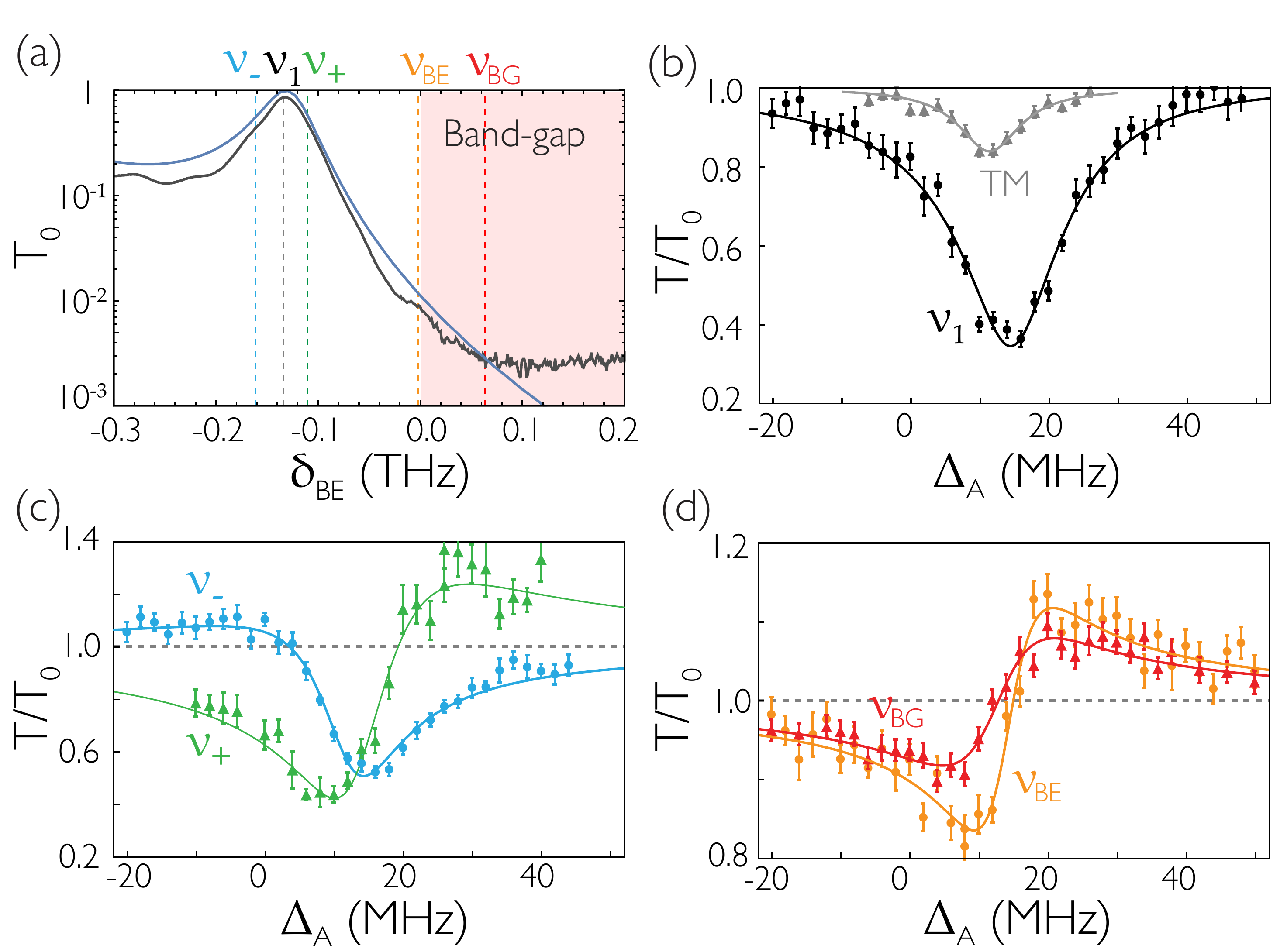}
\caption{Transmission spectra of the PCW without (a) and with trapped atoms (b-d). {\bf (a)} Measured (black) and FDTD simulated (blue) transmission spectra of the PCW without atoms as a function of the probe detuning from the band edge frequency, $\delta_{\rm BE}=\nu_{\rm p}-\nu_{\text{BE}}$.  There is a minimum extinction of 25~dB for the transmitted signal due to fabrication imperfections. {\bf (b-d)} Transmission spectrum for $\bar{N} = 3.0 \pm 0.5$ trapped atoms versus probe detuning $\Delta_{\rm A}= \nu_{\rm p} - \nu_{\text{D1}}$, at several frequencies around the band edge. The solid lines are fits  using the transmission model in \eqref{eq:trans}, averaged over atom positions and different atom numbers.    In {\bf (b)}, the Cs $\rm D_1$ line is aligned to the first `cavity' resonance $\nu_1$, resulting in symmetric spectra for both the TE (black) and TM (gray) modes.   The TE spectra in {\bf (c)} are for frequencies $\nu_{-/+}$  on the two sides of the $\nu_1$ resonance.  The TE spectra in {\bf (d)} are taken at the band edge ($\nu_\text{BE}$, circles) and $60$~GHz ($\nu_{\text{BG}}$, triangles) into the bandgap. The asymmetry of the line-shapes in (c) and (d) implies a large ratio of coherent to dissipative interactions.   
}\label{Fig3}
\end{figure}

After the atoms are loaded into the trap, we send a weak 5 ms probe beam $E_{\rm p}$ with frequency $\nu_{\rm p}$ in either the TE or TM guided mode through the PCW and record the transmitted intensity $|t(\nu_{\rm p})\,E_{\rm p}(\nu_{\rm p})|^2$. The probe beam scans near the  Cs $6S_{1/2},~F=3 \rightarrow 6P_{1/2},~F'=4$ transition.  Each experimental cycle runs at a fixed detuning $\Delta_{\rm A}= \nu_{\rm p} - \nu_{\text{D1}}$ relative to the free-space atomic transition frequency $\nu_{\text{D1}}$. We observe little change of signal during the $5$~ms probing time, suggesting that the atom number is approximately constant over this interval. The band edge of the PCW is tuned thermally by shining an external laser onto a corner of the chip, where its light is absorbed by the silicon substrate.  Hence, the Cs $\rm D_1$ line can be aligned to be either outside or inside the bandgap with an uncertainty $\delta \nu \simeq 5$~GHz.  The transmission for each data point is normalized by the transmission with no atoms ($|t_0 E_{\rm p}|^2$), resulting in a measurement of $T/T_0\equiv|t/t_0|^2$.  The logarithm of the measured and simulated transmission spectrum with no atoms $T_0 = |t_0(\nu_{\rm p})|^2$ is shown in Fig.~\ref{Fig3}(a).  

Examples of transmission spectra with atoms are shown in Figs.~\ref{Fig3}(b-d).   Note that the spectra are shifted 12.5~MHz due to both the AC Stark shift of the dipole trap and the modified Lamb shift induced by the non-guided modes of the PCW. Notably, the transmission spectra at the first `cavity' resonance $\nu_1$ exhibit a characteristic Lorentzian `dip', and they become more and more asymmetric as the frequency moves into the bandgap. 

\textbf{Transmission model.}-- 
We have developed a model to extract quantitative values for collective decay rates and frequency shifts from these atomic transmission spectra \cite{AHC16}. While the formalism of waveguide  \cite{SF05} and cavity QED  \cite{GC85} is well suited for describing atoms coupled to uniform waveguides and cavities, it is not general enough to capture the rich physics of atomic interactions in the vicinity of a PCW. Instead, we describe our system by employing a spin model in terms of the classical electromagnetic Green's function, in which the atoms (or `pseudo-spins' $\sigma_{ge}$ for ground and excited state) interact via the emission and re-absorption of guided photons \cite{DKW02,  BW07, DSF10}. 

The electromagnetic Green's function $\mathbf{G}(\mathbf{r},\mathbf{r}_i,\omega )$ is related to the electric field $\mathbf{E}(\mathbf{r}, \omega)$ emitted by  a dipole $\mathbf{p}_i$ oscillating at frequency $\omega$ at position $\mathbf{r}_i$  by $ \mathbf{E}(\mathbf{r}, \omega)  =\mu_0 \omega^2 \mathbf{G}(\mathbf{r},\mathbf{r}_i,\omega ) \cdot \mathbf{p}_i  $  \cite{NH06,BW07}. The dipole moment operator for atom $i$ is decomposed into	$\hat{\mathbf{p}}_i = \mathbf{d}_i \hat{\sigma}^i_{ge} + \mathbf{d}_i^* \hat{\sigma}^i_{eg}$, where $\mathbf{d}_i $ is the dipole matrix element, and where $\hat{\sigma}_{ge}^i=|g\rangle\langle e|$ is the atomic coherence operator between the ground and excited states.  
The spin model describes a system of $N$ atoms coupled to and driven by a guided mode of the PCW.
In the low saturation and steady-state regime,  expectation values for the atomic coherences ($\sigma^i_{ge} = \langle \hat{\sigma}^i_{ge}\rangle$) are described by a linear system of equations \cite{AHC16, SM} 
\begin{equation}\label{sigmas}
\left(\tilde{\Delta}_{\rm A}+\ii \frac{\Gamma'}{2}\right)\ge^i+\sum_{j=1}^N g_{ij}\,\ge^j= -\Omega_i, 
\end{equation}
where $\tilde{\Delta}_{\rm A}\! =2\pi\Delta_{\rm A}=2\pi(\nu_{\rm p} - \nu_{\text{D1}})$ is the detuning between the probe and the atomic angular frequencies, $\Omega_i$ is the classical drive (Rabi frequency) for the $i$'th atom due to the guided mode input field, and  $g_{ij} = J_{\rm 1D}^{ij} + \ii \Gamma_{\rm 1D}^{ij}/2$  where $J_{\text{1D}}^{ij}=\mu_0\omega_{\rm p}^2/\hbar \,\db_i^*\cdot\text{Re}\,\mathbf{G}(\rb_i,\rb_j,\omega_{\rm p})\cdot\db_j$,  and $\Gamma_{\text{1D}}^{ij} =2\mu_0 \,\omega_{\rm p}^2 /\hbar\,\db_i^*\cdot\text{Im}\,\mathbf{G}(\rb_i,\rb_j,\omega_{\rm p})\cdot\db_j $.  Each atom can also decay into non-guided modes, including free-space, with a decay rate $\Gamma'$. The appearance of the real and imaginary parts of the Green's function in the coherent and dissipative terms has the classical analogue that the in-phase and out-of-phase components of a field with respect to an oscillating dipole store time-averaged energy and perform time-averaged work, respectively. Since the first term in \eqref{sigmas} is diagonal, the atomic coherences can be understood in terms of the eigenvalues $\{ \lambda_{\xi} \}$ for $\xi =\{1,\cdots,N\}$ and eigenfunctions of the matrix $\mathfrak{g}$, whose elements are $g_{ij}$.  The real and imaginary parts of $\{ \lambda_{\xi} \}$ correspond to  frequency shifts and guided mode decay rates, respectively, of the collective atomic mode $\xi$. 

The transmission spectrum can be expressed in terms of the eigenvalues of $\mathfrak{g}$ as \cite{SM,AHC16},
\begin{equation}
\frac{t( \tilde{\Delta}_A,N)}{t_0(\tilde{\Delta}_A) }  = \prod_{\xi=1}^N{ \left( \frac{\tilde{\Delta}_A + \ii \Gamma'/2 }{\tilde{\Delta}_A + \ii \Gamma'/2 +  \lambda_{\xi} }  \right) },
\label{eq:tproduct}
\end{equation}
where $t_0(\tilde{\Delta}_A)$ is the transmission without atoms. 
In the case of a single atom, the only eigenvalue  is proportional to the self-Green's function, $\lambda_{\xi} = g_{ii}$, which implies that the transmission spectrum is a direct measurement of the self-Green's function at the atom's position.    For non-interacting atoms, the off-diagonal elements of $\mathfrak{g}$ are zero, and thus the eigenvalues are single-atom quantities, $\lambda_{\xi}  = g_{ii}$ as there is no cooperative response.  

In contrast, for interacting atoms, the off-diagonal elements are non-negligible, and there is a cooperative response.  In particular, for the atomic frequency inside the bandgap of a photonic crystal, the elements $g_{ij}$ are well approximated by \cite{DHH15}
\begin{equation}
g_{ij} = (J_{\rm 1D} +\ii\Gamma_{\rm 1D}/2)\cos( \pi x_i/a) \cos( \pi x_j/a)  e^{-\kappa_x |x_i - x_j|}, 
\label{eq:gbandgap}
\end{equation}
where the cosine factors arise from the Bloch mode and the decay length $\kappa_x^{-1}$ is due to the exponential decay of the field and results in a finite range of interaction.   
For an infinite photonic crystal, $\Gamma_{\rm 1D}=0$, since the light is localized and there is no dissipation through the guided mode. But for a finite PCW of length $L$, the guided mode dissipation $\Gamma_{\rm 1D}\sim e^{-\kappa_x L} $ is finite due to leakage of the mode out of the edges of the structure.

In the limit where the interaction range $1/\kappa_x$ is much larger than the separation $\delta x_{ij} = |x_i - x_j|$ of the atoms, $\kappa_x \, \delta x_{ij} \lesssim \kappa_x \, \Delta x_{\rm A}  \ll 1$, the guided mode input field  couples predominantly to a single collective ``bright'' mode of the system with eigenvalue $\lambda_B = \sum_{i=1}^N g_{ii}  = \sum_{i=1}^N (J_{\rm 1D}^{ii} + \ii\,\Gamma_{\rm 1D}^{ii}/2)$.  Formally, when $\kappa_x = 0$, the matrix  $\mathfrak{g}$ is separable [$g_{ij} = u_i u_j$ with $u_i \propto \cos( \pi x_i/a)$] and therefore only has one non-zero eigenvalue. In this single bright mode approximation, the transmission spectrum is given by 
\begin{equation}\label{eq:trans}
\frac{t(\tilde{\Delta}_{\rm A},N)}{t_0(\tilde{\Delta}_{\rm A})} =\frac{\tilde{\Delta}_{\rm A}+\ii\Gamma'/2}{\left(\tilde{\Delta}_{\rm A}+\sum_{i=1}^N J_{\rm 1D}^{ii}\right)+\ii\left(\Gamma'+\sum_{i=1}^N\Gamma_{\rm 1D}^{ii}\right)/2}.
\end{equation}

We have confirmed numerically that this single `bright mode' picture is valid within the limits of our uncertainties for the range of frequencies of the measured spectra in Fig.~\ref{Fig3}  
In particular, at the largest detuning into the bandgap $\Delta_{\rm BE}=60$~GHz, we have $\kappa_x \, \Delta x_{\rm A} \simeq 0.2$. However, for atomic frequencies further away from the band edge, this approximation eventually breaks down (e.g., at the bandgap center, $\kappa_x \, \Delta x_{\rm A} \simeq 1.5$).

The single bright mode approximation is also valid in conventional cavity QED. The Green's function matrix is then given by  $g_{ij} =  (J_{\rm 1D} +\ii\Gamma_{\rm 1D}/2)\ \cos(k_{\rm c} x_i) \cos(k_{\rm c}  x_j)$, where $k_{\rm c}$  is the wave-vector of the standing-wave cavity. In this case, $J_{\rm 1D}\propto \Delta_c/(1+ \Delta_c^2/\gamma_c^2)$ and $\Gamma_{\rm 1D}\propto\gamma_c/(1+ \Delta_c^2/\gamma_c^2)$, where $\Delta_{\rm c}$ is the detuning from the cavity resonance and $\gamma_c$ is the cavity linewidth.  Importantly, the ratio between the imaginary dissipative coupling rate to the real coherent coupling rate falls off with inverse detuning, $R_{\rm CQED} = \Gamma_{\rm 1D}/J_{\rm 1D}  = \gamma_c/\Delta_c$ for large $\Delta_c$, whereas in a PCW bandgap, the fall off is exponential with detuning from the band edge.

\textbf{Analysis of measured spectra}-- Equation~(\ref{eq:trans}) provides a direct mapping between the observed transmission spectra of Figs.~\ref{Fig3}(b-d) and the electromagnetic Green's function of the PCW. In particular, the line shape is Lorentzian for purely dissipative dynamics ($J^{ii}_{\text{1D}}=0$). This is precisely what occurs at the frequency of the first cavity mode $\nu_1$, as shown by Fig.~\ref{Fig3}(b). When the GM band edge frequency is moved towards the atomic resonance $\nu_{\rm D1}$, the dispersive interactions are switched on, and the transmission line shape becomes asymmetric, displaying a Fano-like resonance \cite{F1961}, as can be observed in Figs.~\ref{Fig3}(c,d). The appearance of an asymmetry in the atomic spectra directly reveals a significant coherent coupling rate $J_{\rm 1D}$, which is evident for frequencies that lie in the bandgap region. 

For all relevant frequencies, the spectra for the TM guided mode are approximately symmetric, as $J_{\rm 1D}^{\rm TM},\Gamma_{\rm 1D}^{\rm TM}\ll\Gamma'$ for this GM polarization. An example of a TM spectrum is shown in the gray curve of  Fig.~\ref{Fig3}(b). Since the TM bandgap is so far detuned, the TM spectra are insensitive to $\Delta_{\rm BE}$ and serve as a calibration signal.  Using a waveguide transmission model, we fit the TM transmission spectra and extract a TM guided mode decay rate of $\Gamma_{\rm 1D}^{\rm TM} = (0.045\pm.01)\, \Gamma_0$.  This rate is $\sim \!30$ times smaller than the TE guided mode decay rate $\Gamma_{\rm 1D}$ at the first resonance $\nu_1$.  The ratio $\Gamma_{\rm 1D}^{\rm TE} /\Gamma_{\rm 1D}^{\rm TM} \approx 30$  is explained well by the expected slow-light and cavity enhancement of the PCW described in Ref.~\cite{GHH15} and supporting material \cite{SM}. From the TM fits, we also measure $\Gamma' =2\pi \times 9.1$~MHz, which, due to inhomogeneous broadening, is larger than value  $\Gamma' = 2\pi \times 5.0$~MHz  from FDTD numerical calculations \cite{SM}. 
While tuning the band edge to move the atomic frequency $\nu_{\rm D1}$ into in the bandgap, TM spectra are measured in order to confirm \textit{in situ} that the average atom number is approximately constant over the course of the measurements of TE spectra.

\begin{figure}[h!]
\includegraphics[width=\columnwidth]{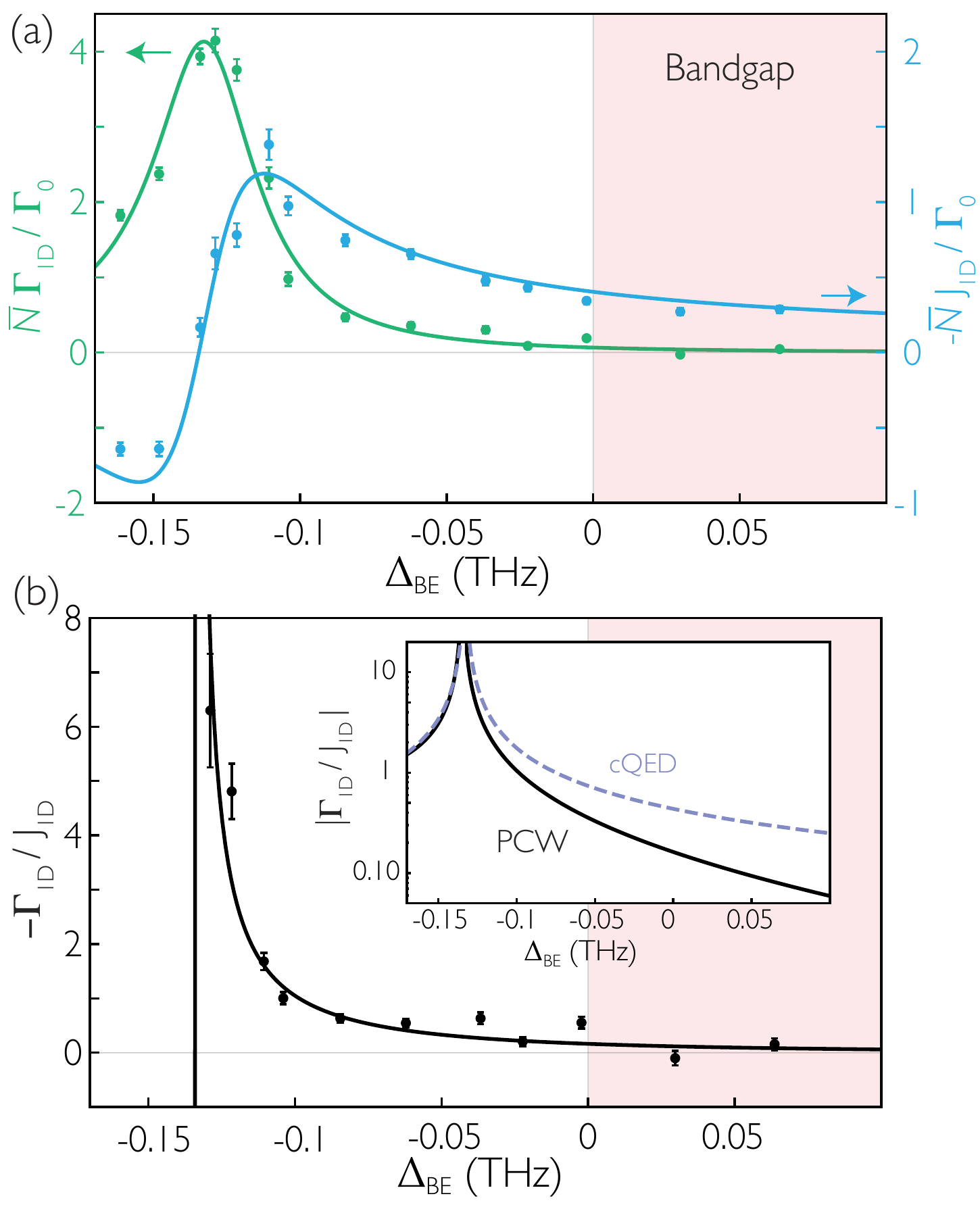}  
\caption{{\bf (a)} Peak dissipative interaction rate $\bar{N} \Gamma_{\text{1D}}$ (green) and coherent rate $\bar{N} J_{\text{1D}}$ (blue) around the band edge. With $\bar{N}$ determined from independent decay rate measurements, the values for $\Gamma_{\text{1D}},J_{\text{1D}}$ are found from fits of the transmission model in Eq. (4) to the measured atomic spectra and are normalized by the free-space decay rate $\Gamma_0 = 2\pi \times 4.56$~MHz for the Cs $\rm D_1$ line.
The lines are the predictions from a numerical model based on 1D transfer matrices 
. {\bf (b)} The measured and calculated ratios $\mathcal{R} = \Gamma_{\text{1D}}/J_{\text{1D}}$.  The average of the two points in the bandgap gives that the ratio of the dissipative to coherent coupling rate is $\mathcal{R} = 0.05 \pm 0.17$.  The inset is a comparison of $\mathcal{R}$ for the PCW calculation (solid) and CQED model (dashed).   From the measured linewidth of the first cavity resonance,  $\gamma_{\rm c} = 60\pm8$~GHz,  CQED predicts that  $\mathcal{R}_{\rm CQED}=\gamma_{\rm c} /\Delta_{\rm c}$, where $\Delta_{\rm c} =  (\nu_{\rm p}- \nu_1)$.  Note that $-J_{\rm 1D}$ is plotted in the figure to more readily compare $\Gamma_{\text{1D}}$ and $J_{\text{1D}}$ as the band edge is approached.}
\label{Fig4}
\end{figure}

To obtain quantitative values for the collective frequency shifts and decay rates by fitting the TE atomic spectra to the spin model, we must account for the fluctuations in atom number and position along the $x$-axis.
As depicted in Fig.~\ref{Fig1}(a) and Fig.~\ref{Fig2}(c), trapped atoms are aproximately free to move along the axis of the device \cite{SM}.
 Their coupling rates are thus modulated by the fast oscillation of the Bloch function, which near the band edge is approximately given by \eqref{eq:gbandgap},  $\Gamma_{\text{1D}}^{ii}(x_i) =  \Gamma_{\text{1D}}\cos^2(x_i \pi/a) $  and  $J_{\text{1D}}^{ii}(x_i) =  J_{\text{1D}}\cos^2(x_i \pi/a) $.  Here $\Gamma_{\text{1D}}$ and $J_{\text{1D}}$ are the peak values. Further, although we know the average atom number $\bar{N} = 3\pm 0.5$ atoms from independent decay-rate measurements \cite{SM}, the atom number for each experiment follows an unknown distribution. 
To model the experimental transmission spectra such as in Fig.~\ref{Fig3}, we average the expression in \eqref{eq:trans} over the atom positions $\{x_i\}$ along the Bloch function and assume a Poisson distribution $P_{\bar{N}}(N)$ for the atom number $N$. We extract peak values $\Gamma_{\text{1D}}$ and $J_{\text{1D}}$, and plot the resulting cooperative rates $\bar{N}\Gamma_{\text{1D}}$ and $\bar{N}J_{\text{1D}}$ in Fig.~\ref{Fig4}(a). In particular, at the first resonance $\nu_1$, the fitted single atom guided-mode decay rate is $\Gamma_{\rm 1D}= (1.4\pm 0.2 )\, \Gamma_0$, which is in good agreement with the decay time measurements, $\Gamma_{\rm 1D}=(1.5\pm 0.2) \, \Gamma_0$ \cite{SM}.  More generally, we find good agreement between our measurements and our model for the transmission, as shown in Fig.~\ref{Fig3}.

The ratio $\mathcal{R}  = \Gamma_\text{1D}/ J_\text{1D}$ is shown in Fig.~\ref{Fig4}(b). Because of the evanescent nature of the field in the bandgap, $\mathcal{R}$ decays exponentially with increasing detuning into the bandgap, $\mathcal{R}\sim e^{-\kappa_x L}$, where $\kappa_x\propto\sqrt{\Delta_{\rm BE}}$ \cite{DHH15}. As displayed in the inset, the ratio between the frequency shift and the GM decay rate diminishes much faster than would be the case in traditional settings such as CQED, for which $\mathcal{R}_{\rm CQED} = \gamma_{\rm c}/\Delta_{\rm c}$, where $\gamma_{\rm c}$ is the cavity linewidth and $\Delta_{\rm c}$ is the detuning from the cavity resonance. Indeed, by performing an average of the last two measured frequencies in the bandgap, we obtain $\mathcal{R}=0.05\pm0.17$,  whereas $\mathcal{R}_{\rm CQED}=0.30\pm0.04$, where we have taken the cavity linewidth to be a value consistent with the linewidth of the first cavity mode of the PCW ($\gamma_{\rm c} = 60\pm8$~GHz).  We can then infer that the ratio of dispersive to dissipative rates for guided mode atom-atom interactions (i.e.,  $1/\mathcal{R}$) is significantly larger than is the case in conventional optical physics (e.g., CQED). 

Beyond the detailed modeling involving \eqref{eq:trans} averaged over fluctuations in atom number and position, we also fit the spectra with a generic transmission model with no averaging, as shown in the supporting material \cite{SM}. We find that the effective values for the guided mode decay rate and frequency shift are related to $\bar{N}\Gamma_{\rm 1D}$ and $\bar{N}J_{\rm 1D}$ in Fig.~\ref{Fig4}(a) by a simple scale factor related to the averaging of the Bloch function $\cos^2(\pi x/a)$.

Despite favorable scaling between the collective frequency shifts and the guided mode decay rates, there is still one obstacle to overcome towards purely dispersive atomic interactions, namely atomic emission into non-guided modes (characterized by $\Gamma'$). For the current PCW structure, the FDTD simulated value of this decay rate is $\Gamma'\simeq1.1\,\Gamma_0$ \cite{GHH15} for the relevant frequencies of our experiment. Fortunately, it has been shown that suitable engineering of a wide variety of nanophotonic structures can lead to significant reductions in $\Gamma'/\Gamma_0$ \cite{Hung13}. For example, Ref.~\cite{LMS15} reviews possibilities to achieve $\Gamma' \simeq0.1 \Gamma_0$.

\textit{Concluding remarks and outlook}-- In conclusion, we report the first observation of cooperative atom interactions in the bandgap of a photonic crystal waveguide. By tuning the band edge frequency of the photonic crystal waveguide, we are able to modify the interactions between the atoms that are trapped close to the device, reducing the dissipative relative to coherent coupling for frequencies inside the bandgap of the PCW. Equipped with a theoretical model based on the electromagnetic Green's function of the alligator photonic crystal waveguide, we infer quantitative values for the collective frequency shifts and decay rates experienced by the atoms. Moreover, we infer a suppression of the dissipative interactions with respect to the coherent ones several times larger than is customarily obtained in AMO physics. This measurement provides the first stepping stone towards the realization of quantum many body physics in bandgap systems.

Moreover, near-term extensions of our experiment open the door to exploring new physical scenarios by employing atoms coupled to PCWs. By trapping the atoms at the center of the device with guided modes \cite{Hung13}, we expect a 6-fold increase to both coupling strengths $J_{\rm 1D}$ and $\Gamma_{\rm 1D}$ relative to $\Gamma'$.  Moreover, by probing the atoms with the Cs $D_2$ line tuned to the upper band edge, where the intensity at the position of the atoms is larger, we expect a further improvement by a factor of two.  Combining these two effects,  we expect a significant enhancement of interactions via guided modes as compared to conventional free space interactions, namely $J_{\rm 1D},\Gamma_{\rm 1D} > 10\times \Gamma^{\prime}$. This could enable investigations of new paradigms for atom-photon interactions, such as the recently proposed multi-photon dressed states \cite{calajo,shi}.

\textbf{Note added}-- After the submission of this manuscript, Ref.~\cite{LH16} reported measurements of transmission spectra for a  superconducting qubit placed within the bandgap of a microwave photonic crystal.

\textit{Acknowledgments}--We gratefully acknowledge the contributions of O. J. Painter and his group, including for fabrication and clean-room facilities. We further acknowledge A. Burgers, C.-L. Hung, J. Laurat, M. J. Martin, A. C. McClung,  J. A. Muniz, and L. Peng. Funding is provided by the DOD NSSEFF program, the AFOSR QuMPASS MURI, NSF Grant PHY-1205729 the ONR QOMAND MURI, and the IQIM, an NSF Physics Frontiers Center with support of the Moore Foundation. A. G. was supported by the Nakajima Foundation. A. A.-G. and M. L. were supported by the IQIM Postdoctoral Fellowship. A. A.-G. also acknowledges support from the Global Marie Curie Fellowship LANTERN (655701). S.-P. Y. acknowledges support from the International Fulbright Science and Technology Award. DEC acknowledges support from Fundacio Privada Cellex Barcelona, Marie Curie CIG ATOMNANO, MINECO Severo Ochoa Grant SEV-2015-0522, and ERC Starting Grant FoQAL.

\pagebreak
\widetext
\begin{center}
\textbf{\large Supplemental Information: Atom-atom interactions around the band edge of a photonic crystal waveguide}
\end{center}
\setcounter{equation}{0}
\setcounter{figure}{0}
\setcounter{table}{0}
\setcounter{page}{1}
\makeatletter
\renewcommand{\theequation}{S\arabic{equation}}
\renewcommand{\thefigure}{S\arabic{figure}}
\renewcommand{\bibnumfmt}[1]{[S#1]}
\renewcommand{\citenumfont}[1]{S#1}

\section{Introduction}

In our results of the main text, we measure collective frequency shifts and decay rates for atoms trapped near a photonic crystal waveguide (PCW).  In our previous work in Ref.~\cite{GHH15}, we  trapped multiple atoms in an optical dipole-force trap above the PCW. We operated with the atomic frequency outside the bandgap in a regime with large decay rate $\Gamma_{\rm 1D}$ and small coherent coupling rate $J_{\rm 1D}$.  By varying the density and observing the superradiant decay of the atoms $\Gamma_{\rm tot}^{(\bar{N})} = \Gamma_{\rm SR}(\bar{N})+ \Gamma_{\rm 1D} + \Gamma'$, we inferred the single-atom guided-mode decay rate $\Gamma_{\rm 1D}$ and the average number of atoms $\bar{N}$. Importantly, this measured single-atom decay rate $\Gamma_{\rm 1D}$ agreed well with the finite-difference time-domain (FDTD) simulations at the calculated trap location. This good agreement is in part due to the nanometer-scale accuracy in which the alligator PCWs are fabricated, which is required for both the band-edge alignment and the device quality. 

In our current manuscript, the band-edge of the PCW is tuned around the resonance frequency of the atoms, and we observe the dominance of the  guided-mode coherent coupling rates $J_{\rm 1D}$ over the dissipative coupling rates $\Gamma_{\rm 1D}$, which is associated with atomic radiative processes for operation within the bandgap. To extract quantitative values for these parameters from our measurements of transmission spectra for atoms trapped along a PCW, we have developed theoretical techniques based upon Green's functions for the PCW, which are new to atomic physics. As in Ref. \cite{GHH15}, the average number of atoms $\bar{N}$ is measured by way of transient decay. Our principal finding relates to the turning-off of the guided-mode decay rate $\Gamma_{\rm 1D}$, which in the bandgap is predicted to be exponentially suppressed, while nonetheless, retaining appreciable coherent processes described by $J_{\rm 1D}$. 

For the spectra in our current manuscript, the transmission through the device decreases exponentially in the bandgap, and more time is required to measure the transmission spectra as compared to our work in Ref. \cite{SGHH15}.   Unfortunately, cesium slowly coats the PCW during the measurement, both degrading the device quality and shifting the band-edge out of the thermal tuning range.  As a result, each device only has a limited lifetime for making transmission measurements.  For our current experiment, we first repeated superradiance measurements outside the bandgap at the first resonance $\nu_1$ of the PCW in order to determine the average number of atoms $\bar{N}$ and the single-atom guided-mode decay rate $\Gamma_{\rm 1D}$, and to show that the atoms behave as a collective emitter.  Then, with an average number of $\bar{N}\simeq3$,  we measured transmission spectra as the atomic frequency is shifted into the bandgap.  We simultaneously measured the TM spectra to verify that the atom number is constant over the course of the measurements of the TE spectra.  

\section{Alligator photonic crystal waveguide design and fabrication}\label{sec:alligator}
\begin{figure*}[h!]  
  \includegraphics[width=0.8\linewidth]{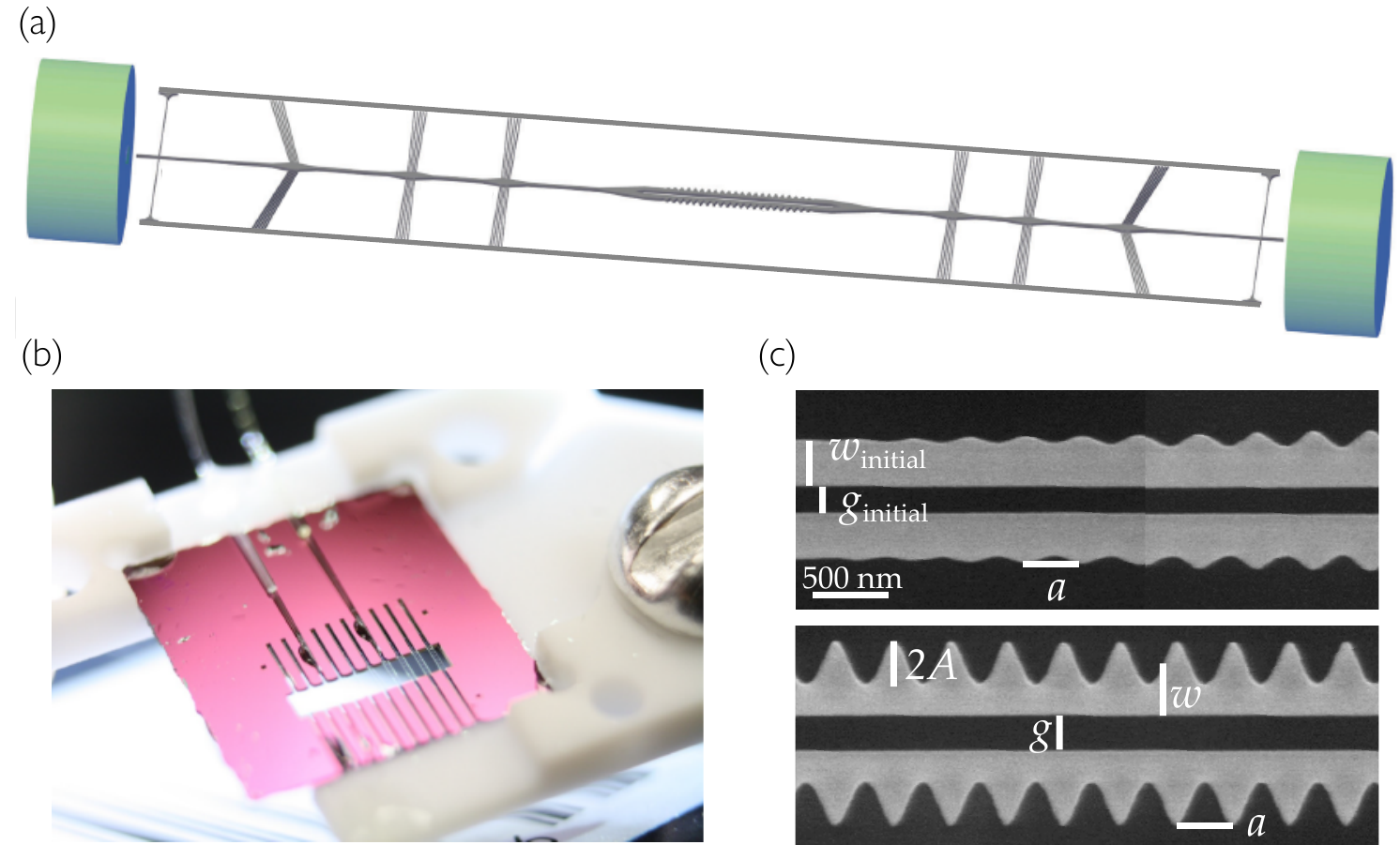}
\caption[]{ Alligator photonic crystal waveguide (PCW) chip and device overview, taken from Ref.~\cite{SYHM14}. {\bf(a)}  Schematic of the entire device.  The alligator photonic crystal waveguide (PCW) is at the center.  Optical fibers (green) on both ends couple light into and out of the waveguide.  The waveguide is surrounded by supporting and cooling structures. 
{\bf(b)}  Image of a $10\times10$~mm PCW chip, taken from Ref.~\cite{SYHM14}.  Multiple waveguides stretch across the window of the chip, with the PCWs at the center of the window.  The window provides optical access for trapping and cooling atoms around the device. 
{\bf(c)}  Overview of device variables.  The lattice constant for the entire device is $a=370$~nm.  The device dimensions are measured with an SEM and are calibrated to the lattice constant. The device dimensions are $w = 310\pm 10$~nm,  $2A = 262 \pm 10$ nm,  $g= 220\pm 10$ nm, $w_{\rm initial} = 268\pm 15$ nm, $g_{\rm initial} = 165\pm 10$ nm.  The thickness of the silicon nitride is $185\pm5$ nm. The index of refraction for Si$_3$N$_4$ is $n=2.0$ around our frequencies of interest.      
    } 
\label{fig:alligator}
\end{figure*}

The schematic of the device is shown in Fig.~\ref{fig:alligator}(a).    Light is coupled into and out of the device by mode-matching the output of an optical fiber to that of a terminated rectangular-shaped waveguide on both sides of the device \cite{SYHM14}.  The fibers are glued permanently in etched v-grooves at optimized coupling positions.  The design and fabrication of the alligator photonic crystal waveguide (PCW) are detailed in Ref.~\cite{SYHM14}.  The PCW is fabricated on a 200 $\mu$m silicon (Si) chip coated with a 200 nm thick silicon nitride (SiN) film.  The SiN device is suspended across a $2$-mm-wide window after the silicon substrate beneath it is removed,  as shown in the image of Fig.~\ref{fig:alligator}(b).  The window allows optical access for the trapping and cooling of atoms around the device.   

The dielectric TE mode band edge ($\nu_{\rm BE}$) is aligned to within $200$~GHz of the Cs D1 line ($\nu_{\rm D1}=335.12$~THz) via a low-power inductively-coupled reactive-ion CF$_4$ etch.  The directional etch thins the SiN layer at a rate of 3 nm/min until a transmission measurement confirms alignment of the band edge.  The final geometric dimensions of the device used in the main text are given in Fig.~\ref{fig:alligator}(c).

\begin{figure*}[ht]  
\centering
  \includegraphics[width=0.8\linewidth]{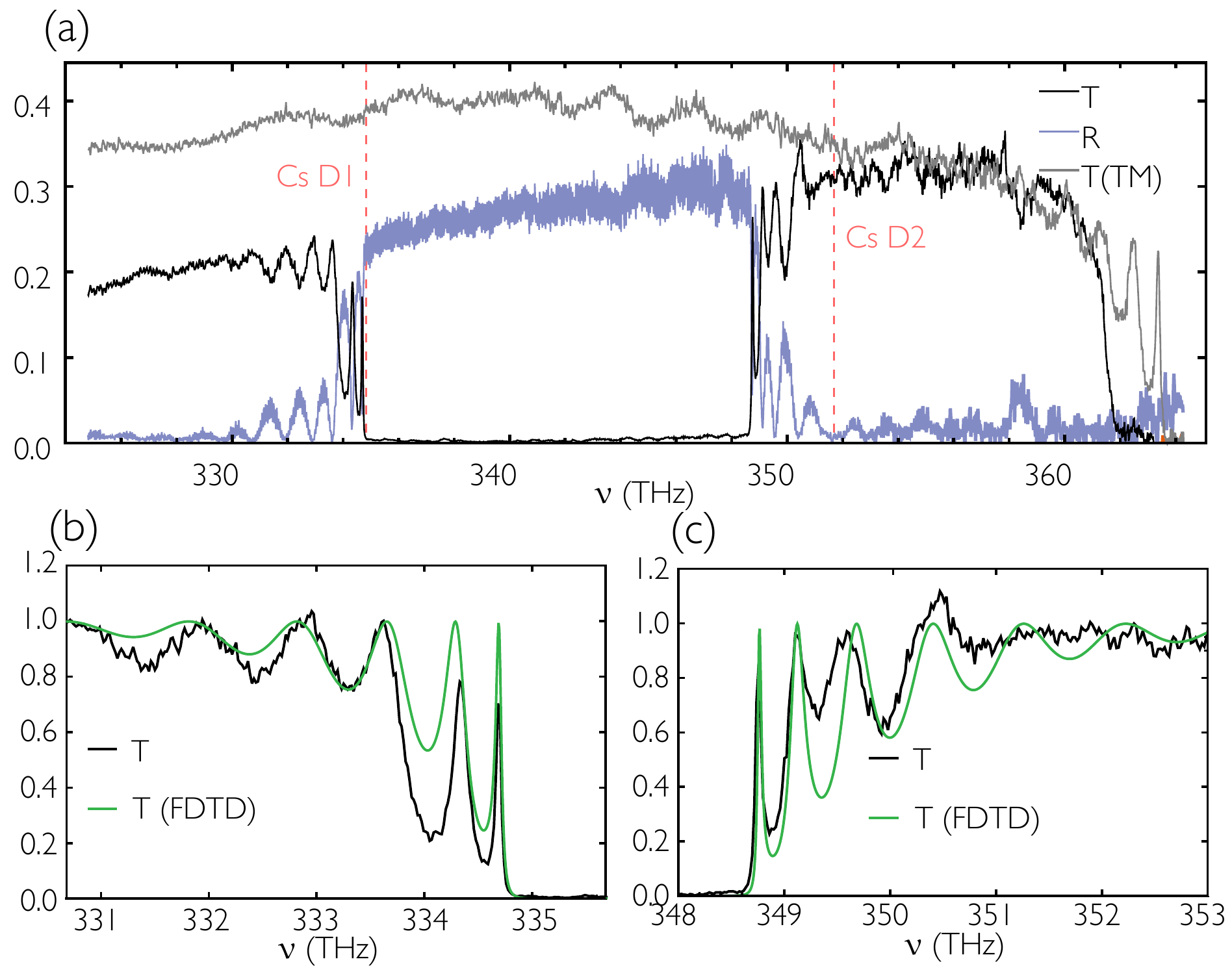}   
	\caption[]{ Measured and simulated transmission and reflection spectra. 
	{\bf(a)}  Transmission (black) and reflection (blue) spectra through the entire chip for the TE mode (polarization in the plane of the device).  The red dashed lines are the Cs D$_1$ (335.1 THz) and D$_2$ (351.7 THz) lines.   The TE transmission efficiency through the entire device near the dielectric band edge is $\sim 23\%$, indicating that the single pass efficiency from the fiber to device is approximately $49\%$.   Most of the loss is due to the waveguide-to-fiber coupling section. The gray line is the TM transmission (polarization perpendicular to the plane of the device).  Note that the lower band edge of the TM mode is visible at around $365$~THz, but is far detuned from both Cs D$_{1,2}$ transitions.  
 {\bf(b-c)}  TE transmission data is normalized and compared to a finite-difference time-domain (FDTD) simulation~\cite{SLumerical}.  The simulation uses the measured device parameters in Fig.~\ref{fig:alligator}, but adjusted within the uncertainty of the measurements so that the position of the first resonances match those in the measured spectra.}  \label{fig:TR}
\end{figure*}

For the experiment, the chip is placed at the center of a ultra-high vacuum chamber, and the optical fibers exit through Teflon fiber feed-throughs.  We measure the transmission through a device using a  super luminescent  diode (SLD) as the source and an optical spectrum analyzer (OSA) as the detector.  The measured transmission and reflection spectra are shown in Fig.~\ref{fig:TR}(a).  The transmission spectra near the lower (dielectric) and upper (air) band edge are compared to an FDTD simulation in Fig.~\ref{fig:TR}(b-c).  

\section{Alligator dispersion relation from scattering images }\label{sec:scattered}

\begin{figure*}[ht]   
\centering
  \includegraphics[width=0.5\linewidth]{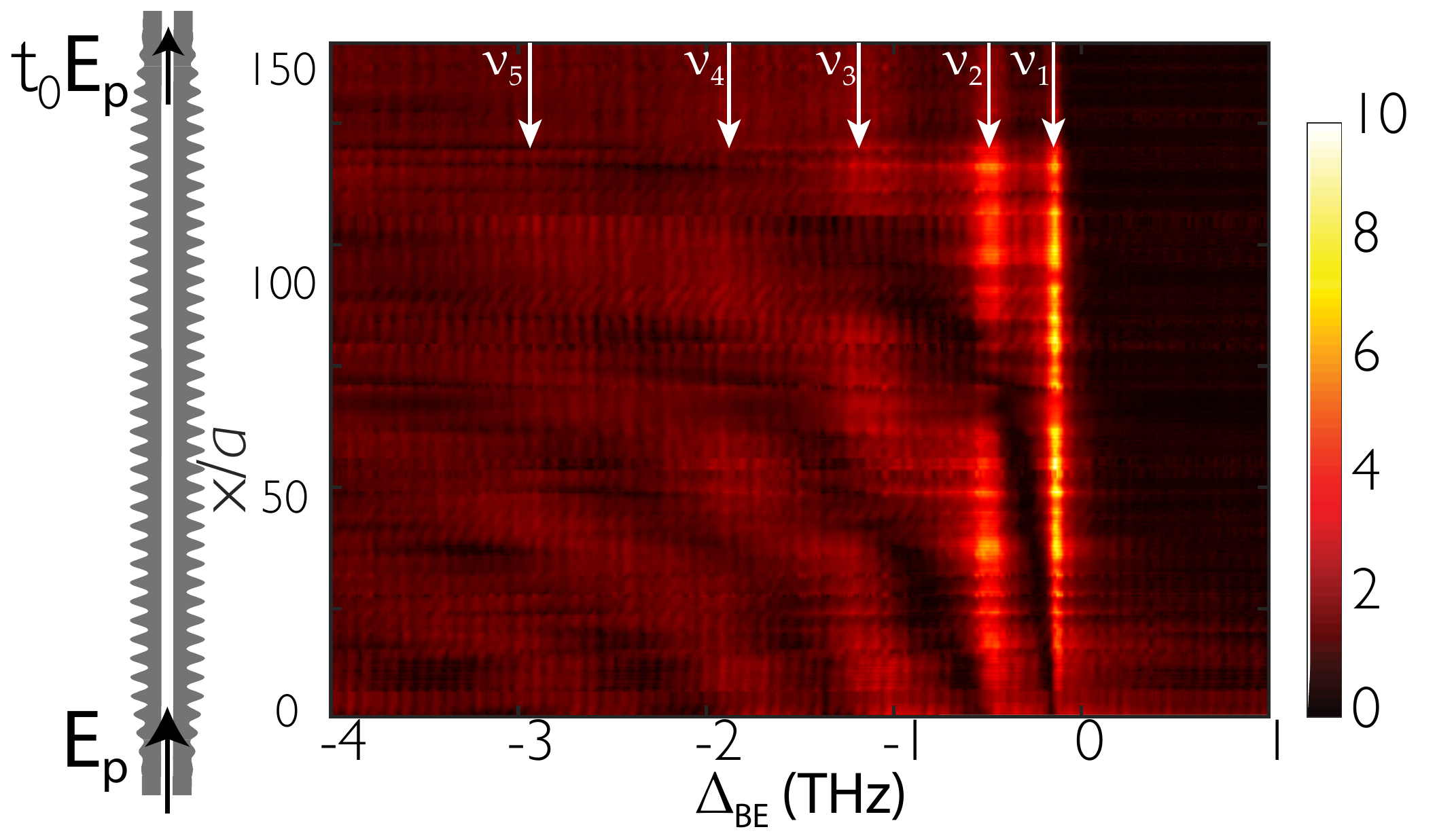}   
	\caption[]{   Normalized magnitude of the scattered electric field  of the PCW for frequencies  $\Delta_{\rm BE} = \nu_p - \nu_{\rm BE}$ around the band edge. The schematic on the left shows the PCW with the number of unit cells reduced by 5.}    \label{fig:Scattered}  
\end{figure*}

Here, we describe the analysis performed for the PCW dispersion relations in Fig.~2(e) of the manuscript.  We send a single-frequency laser beam through the device and image the scattered light with a microscope.  We integrate the image over the width of the PCW to produce a single plot of intensity versus position.  Then we scan the laser frequency around the lower band edge to produce a 2D plot of scattered intensity as a function of position $x$ along the device and frequency $\nu$ of the input light.  

The weak scattered light comes from small fabrication imperfections or intrinsic material defects and serves as a probe of the local intensity. Since each scatterer emits light at a different rate, we have to normalize the scattered light by a reference intensity spectrum in which the intensity of the device is known.  For this reference spectrum, we average over the intensities for frequencies far from the band edge, where the PCW behaves like a waveguide, and where the local intensity in the device is approximately  constant.   The normalized data is shown in Fig.~\ref{fig:Scattered}, and a zoomed-in version is in Fig.~2(a) of the manuscript.  

In the FDTD simulation described above, we calculate the intensity along the center of the device for frequencies around the band edge.  Taking the maximum intensity in each unit cell and normalizing by the intensity in the waveguide regime, we produce Fig.~2(b) in the main text.    

Next, we fit the intensity spectrum at a given frequency to a model in order to extract the wave-vector for that frequency. Near the band edge, the field in an infinite PCW is well approximated by $E(x) \propto \cos(x \pi/a) e^{\ii \delta k_x  x}$, where  $\delta k_x = \pi/a - k_x $ in the propagating band $(\Delta_{\rm  BE} < 0 )$ and $\delta k_x = \ii \kappa_x $ inside the bandgap ($\Delta_{\rm BE} > 0 $) 
The edges of a finite photonic crystal reflect with $R_t$ due to a large group index mismatch between the waveguide section and the photonic crystal waveguide.  The resonances of the weak cavity result in the cavity-like intensity profiles seen at frequencies $\nu_{1,2,3,4,5}$ in Fig.~\ref{fig:Scattered}.   The intensity at a point $x$ along a finite photonic crystal of length $L$ is well approximated by a model based on the intensity in a cavity with two mirrors of reflectivity $R_{\rm t}$,
\begin{equation}
|E(x)|^2= I_1\,|e^{\ii \delta k_x x} - R_{\rm t} e^{2 \ii \delta k_x L} e^{-\ii \delta k_x x}|^2,  
\label{eq:Imodel}
\end{equation}
where $I_1$ is related to the overall intensity. This expression ignores the fast oscillations of the Bloch function, which go as $\cos^2(x \pi/a)$.   Note that in the bandgap (when $\kappa_x L \gg 1$), the intensity model reduces to an exponential decay: $|E(x)|^2 \approx I_1\,e^{-2 \kappa_x x} $.     Interestingly, at the band edge ($\delta k_x \rightarrow 0$, $R_t \rightarrow 1$), the intensity displays a quadratic dependence on the position, $|E(x)|^2 \propto (L-x)^2 $. 

\begin{figure*}[ht]  
\centering
  \includegraphics[width=0.8\linewidth]{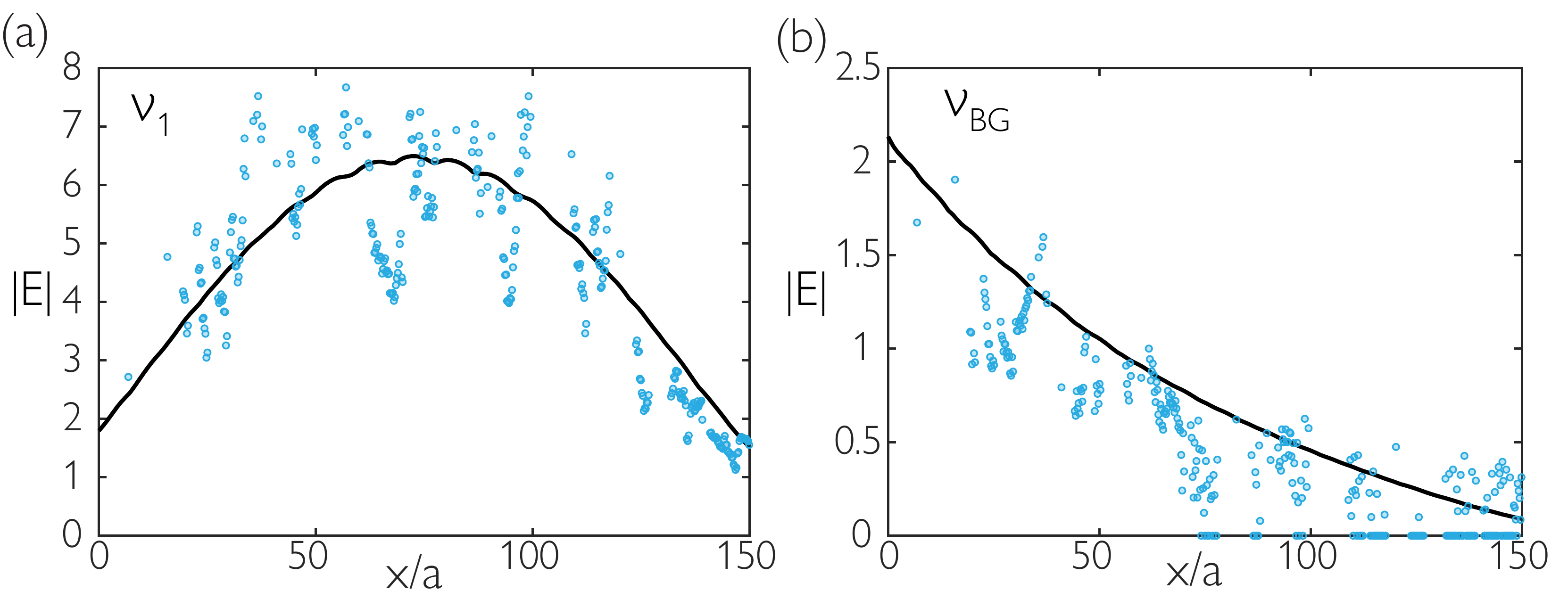}   
	\caption[]{  The electric field magnitude in the PCW at the first resonance $\nu_1$ {\bf (a)}, and in the bandgap $\nu_{\rm BG}= \nu_{\rm BE} + 60$~GHz  {\bf (b)}.   The points show measured data, and the black lines are from an FDTD simulation.  The electric field magnitude $|E|$ is normalized by the electric field magnitude far from the band edge; thus, these plots gives the enhancement of $|E|$ over the waveguide regime.}  \label{fig:fields}
\end{figure*}

For each frequency, we fit the intensity along the nominal cells with \eqref{eq:Imodel} and extract $\delta k_x $.   This procedure allows us to map out the dispersion relation $\delta k_x(\Delta_{\rm BE}) $, which we show in Fig.~2(e) for the measured and simulated data.  From the simulated fits, we find that the effective length of the cavity is $162$ cells, which is slightly longer than the $150$ nominal cells.  This is expected since the cavity field can leak into the tapering sections. We use this length for the fits of the measured data.  Examples of the measured and simulated intensity are shown in Fig.~\ref{fig:fields}.  The fluctuation of the intensity, even after the normalization, is most likely due to the spatial profile of Bloch mode.   The normalization trace is taken by averaging data for excitation frequencies further away from the band-edge where the Bloch mode contrast is reduced, whereas the data closer to the band-edge has a large Bloch mode fringe visibility.  However, the fluctuations do not affect the statistical fits at the level of accuracy required for the dispersion relation in our current work.

The frequency for which $\delta k_x = 0 $ is defined as the band edge frequency $\nu_{\rm BE}$.  To extract this frequency and the curvature of the dispersion relation near the band edge, we fit the measured and simulated dispersion relations with a dispersion model \cite{SGHH15},
\begin{equation}  
\delta k_x(\nu) = \frac{2 \pi}{a} \sqrt{ \frac{(\nu_{\rm BE2}-\nu)(\nu_{\rm BE}-\nu) }{ 4 \zeta^2 - (\nu_{\rm BE2} - \nu_{\rm BE})^2 }}, 
\label{eq:VH}
\end{equation}
where $\nu_{\rm BE}$ ($\nu_{\rm BE2}$) is the lower (upper) band edge frequency, and $\zeta$ is a frequency related to the curvature of the band near the band edge. 
From the measured data fits, the distance between the first resonance and band edge is $\nu_{\rm BE}-\nu_{1}  = 133 \pm 9 $~GHz and $\zeta = 227\pm3$~THz. 
  The simulated data give $\nu_{\rm BE}-\nu_{1} = 135.0$~GHz and the curvature parameter is $\zeta = 226.0 $~THz.   These values are in good agreement with the dispersion relation from the eigenmode simulation of the infinite PCW in Fig.~1(c) of the main text, which  gives $\zeta = 229.1$~THz.
  
\section{Side-illumination trap} 
\begin{figure*}[ht]   
\centering
\includegraphics[width=0.8\linewidth]{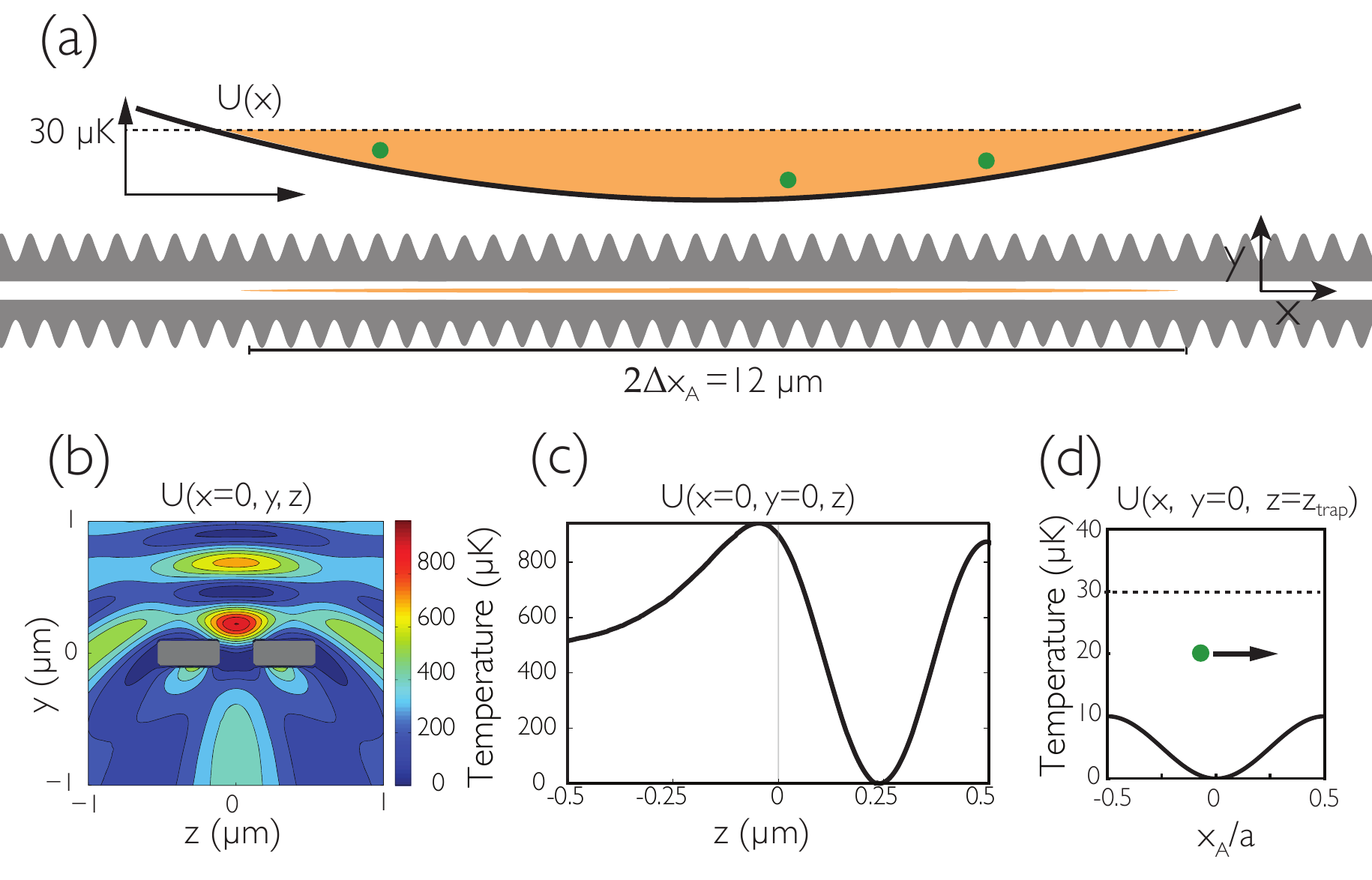} 
\caption{{\bf (a)} Schematic of the atoms in the side-illumination (SI) trap. Given the estimated atom temperature of $~30$~$\mu$K, we infer that the atoms are confined to a length of $2 \Delta x_A = 12$~$\mu$m along the $x$-axis.   {\bf (b,c,d)} FORT potentials for the SI trap simulation (b) in the $y$-$z$ plane \cite{SGHH15}, (c) along the $z$-axis, and (d) along the $x$-axis. 
}
 \label{fig:trap}
\end{figure*}

In Fig.~\ref{fig:trap}(a), we show a schematic of the side-illumination (SI) trap.  The side-illumination beam is nearly perpendicular to the axis of the device, has a 50~$\mu$m diameter, and has a polarization aligned to the axis of the device.   The orange areas represent the approximate localization of the atoms along $x,y$. By time-of-flight measurements of atoms in the dipole traps, we estimate an atomic `temperature' of approximately 30~$\mu$K.  From the beam waist and atom temperature, we can infer that the atoms are localized to $2 \Delta x_A = 12$~$\mu$m along the $x$-axis.

Simulations of the FORT potential for the SI trap are shown in Fig.~\ref{fig:trap}(b-d).  The simulations are performed for the infinite structure with COMSOL. The trap depth is calibrated with the $12$~MHz AC Stark shift measured from the atomic spectra.  Figure~\ref{fig:trap}(b) shows the trap potential in the $y$-$z$ plane.  Atoms that are significantly hotter than $\sim 100$~$\mu$K are expected to crash into the device along the diagonal directions due to Casimir-Polder forces.  Figure~\ref{fig:trap}(c) shows the trapping potential along the $z$-axis.  Atoms are trapped at $z=240$~nm.  
Figure~\ref{fig:trap}(d) shows the trap along the $x$-axis. Due to the photonic crystal, the trap modulates by $\sim 10$~$\mu$K along the $x$-axis, but this is  significantly smaller than the estimated trap temperature.

In addition to the results in Fig.~\ref{fig:trap}, we have also carried out numerical modeling of the optical trap using Lumerical simulations \cite{SLumerical} of the actual finite length PCW and tapers shown in Fig.~\ref{fig:alligator}. We have as well included Casimir-Polder potentials as in Ref. \cite{SHung13}. More details of the trap are discussed in Ref.~\cite{SGHH15}.

\section{Transmission model and atomic spectra fits}
Here we give a more detailed description of the transmission model in the main text, which follows the derivation given in Ref.~\cite{SAHC16}.  A system of $N$ atoms coupled to a radiation field can be described using formalism based on the classical Green's function~\cite{SBW07,SDKW02}. In the Markovian limit, the field can be eliminated to obtain a master equation that describes the interactions between the atoms,   $\dot{\hat{\rho}}_{\rm A} = -\frac{\ii}{\hbar}[H,\hat{\rho}_{\rm A}]+\mathcal{L}[\hat{\rho}_{\rm A}]$.   Here the Hamiltonian $H$ gives the coherent evolution of the system, 
 \begin{equation}
H=-\hbar\sum_{j=1}^N \tilde{\Delta}_{\rm A}\hat{\sigma}_{ee}^j  -\hbar \sum_{j,i=1}^N J^{ji}_{\text{1D}}\hat{\sigma}_{eg}^j\hat{\sigma}_{ge}^i -
\hbar \sum_{j=1}^N \left(\Omega_j\hat{\sigma}_{eg}^j+ \Omega_j^*\hat{\sigma}_{ge}^j\right),
\end{equation}
and the Lindblad operator $\mathcal{L}[\hat{\rho}_{\rm A}]$ gives the dissipation of the system,
\begin{align}
\mathcal{L}[\rho_{\rm A}]=&\sum_{j,i=1}^N\frac{\Gamma' \delta_{ji}+\Gamma_{\text{1D}}^{ji}}{2}\\\nonumber
&\times\left(2\hat{\sigma}_{ge}^j\hat{\rho}_{\rm A}\hat{\sigma}_{eg}^i-\hat{\sigma}_{eg}^j\hat{\sigma}_{ge}^i\hat{\rho}_{\rm A}-\hat{\rho}_{\rm A}\hat{\sigma}_{eg}^j\hat{\sigma}_{ge}^i\right).
\end{align}
The Hamiltonian and Lindblad are expressed in terms of the atomic coherence operator $\hat{\sigma}_{ge}^j=|g\rangle\langle e|$ between the ground and excited states of atom $j$.  
The Hamiltonian contains terms for the free-atom evolution, the coherent atom-atom interactions, and the classical drive, respectively.  $\tilde{\Delta}_{\rm A}\! =2\pi\Delta_{\rm A}=2\pi(\nu_{\rm p} - \nu_{\text{D1}})$ is the detuning between the probe and the atomic angular frequencies. $\Omega_j$ is the Rabi frequency for atom $j$ due to the guided-mode field.  
The atom-atom spin-exchange rate $J_{\rm 1D}^{ji}$ is expressed in terms of the real part of the guided mode Green's function as 
\begin{equation}\label{eq:Jdef}
J_{\text{1D}}^{ji}=(\mu_0\omega_{\rm p}^2/\hbar)\,\db_j^*\cdot\text{Re}\,\mathbf{G}(\rb_j,\rb_i,\omega_{\rm p})\cdot\db_i,
\end{equation}
where $\omega_{\rm p}=2\pi\nu_{\rm p}$ and $\textbf{d}_j$ is the dipole matrix element of atom $j$.   
The Lindblad term is responsible for the dissipative interactions in the system, which include atomic decay into non-guided ($\Gamma'$) and guided ($\Gamma_{\text{1D}}^{ji}$) modes.
The decay rate into the guided mode is written in terms of the imaginary part of the Green's function as 
\begin{equation} \label{eq:Gammadef}
\Gamma_{\text{1D}}^{ji} =2(\mu_0\omega_{\rm p}^2/\hbar)\,\db_j^*\cdot\text{Im}\,\mathbf{G}(\rb_j,\rb_i,\omega_{\rm p})\cdot\db_i.
\end{equation}
For low atomic density along the PCW, the non-guided emission rate $\Gamma'$ is not cooperative, and is described here as a single-atom effect, with $\delta_{ji}$ as the Kronecker delta.

In the low saturation regime, the Heisenberg equations for the expectation value of the atomic coherences ($\braket{\hat{\sigma}_{eg}}=\eg$) can be solved for with the master equation leading to
\begin{equation}\label{sigmas}
\dot{\sigma}_{ge}^j=\ii\left(\tilde{\Delta}_{\rm A}+\ii\frac{\Gamma'}{2}\right)\ge^j+\ii \, \Omega_j
+\ii\sum_{i=1}^N g_{ji}\,\ge^i,
\end{equation}
where the complex coupling rate is
\begin{equation}  \label{eq:g}
g_{ij}=J_{\text{1D}}^{ij} +\ii \Gamma_{\text{1D}}^{ij}/2 = (\mu_0\omega_{\rm p}^2/\hbar)\,\db_i^*\cdot \mathbf{G}(\rb_i,\rb_j,\omega_{\rm p})\cdot\db_j , 
\end{equation}
which is the Green's function between atoms $i$ and $j$ projected onto the respective dipole matrix elements.  In the steady-state solution, the time derivative is set to zero and result is the linear system of equations for the atomic coherences given in the main text.

The electric field in the system can be expressed in terms of the  input probe field $\mathbf{E}^+(\mathbf{r}, \omega_p)$ and solutions for the atomic coherences  \cite{SAHC16},
\begin{equation}\label{eq:field}
{\mathbf{E}}^+(\mathbf{r},\omega_{\rm p})={\mathbf{E}}^+_{\rm p}( \mathbf{r},\omega_{\rm p}) +\mu_0\omega_{\rm p}^2\sum_j\mathbf{G}(\mathbf{r},\mathbf{r}_j,\omega_{\rm p})\cdot\mathbf{d}_j {\sigma}^j_{ge}.
\end{equation}
An expression for the transmission through a quasi-1D structure  can be derived by solving the steady state system of equations in \eqref{sigmas} for the  atomic coherences $\sigma_{ge}^{j} $ and substituting them into \eqref{eq:field}.  The expression can then be simplified in the case where the dipole moments are real, in which case $\mathfrak{g}$ is a complex symmetric matrix with eigenvectors and eigenvalues $\mathfrak{g} \, \mathbf{u}_{\xi}  = \lambda_{\xi} \, \mathbf{u}_{\xi}$, and when the Green's function is well represented by a 1D Green's function. The final result is \cite{SAHC16},
\begin{equation}
\frac{t( \tilde{\Delta}_A,N)}{t_0(\tilde{\Delta}_A) }  = \prod_{\xi=1}^N{ \left( \frac{\tilde{\Delta}_A + \ii \Gamma'/2 }{\tilde{\Delta}_A + \ii \Gamma'/2 +  \lambda_{\xi} }  \right) },
\label{eq:tproduct}
\end{equation}
where $t_0(\tilde{\Delta}_A)$ is the transmission without atoms.   

In the bandgap, the matrix  $\mathfrak{g}$ of elements $g_{ij}$ is well approximated by 
\begin{equation}   \label{eq:gbandgap}
g_{ij} = (J_{\rm 1D} +\ii\Gamma_{\rm 1D}/2)\cos( \pi x_i/a) \cos( \pi x_j/a)  e^{-\kappa_x |x_i - x_j|}.  
\end{equation}
As  discussed in the main text, when the interaction range $1/\kappa_x$ is much larger than the separation distance ($\kappa_x |x_i-x_j|\ll 1 $), there is only a single atomic `bright mode', for which the frequency shift and guided-mode decay rate are given by $ \sum_{i=1}^N J_{\rm 1D}^{ii} $ and   $ \sum_{i=1}^N \Gamma_{\rm 1D}^{ii}$. The transmission spectra for $N$ atoms in the `single-bright-mode' approximation is given by  
\begin{equation}  
T( \Delta_{\rm A} , N) = T_0(\Delta_{\rm A}) \left|  \frac{\tilde{\Delta}_{\rm A} + \ii \Gamma'/2}{\tilde{\Delta}_{\rm A} + \ii \Gamma'/2 + \sum_i( J_{\rm 1D}^{ii} +  \ii \Gamma_{\rm 1D}^{ii}/2  )}  \right|^2,
\label{eq:T}
\end{equation}
where $\tilde{\Delta}_{\rm A}=2\pi\Delta_{\rm A}=2\pi(\nu_p - \nu_{D1})$ is the detuning between the pump and the atomic frequency, and $T_0(\Delta_{\rm A})$ is the device transmission when no atoms are present.   

Explicitly accounting for the atoms' positions by substituting \eqref{eq:gbandgap} into \eqref{eq:T},  the transmission is given by
\begin{align}  \label{eq:Tsb}
&T( \Delta_{\rm A} , N; x_1,...,x_N) / T_0(\Delta_{\rm A}) =\\\nonumber
&\left|  \frac{\Delta_{\rm A}' + \ii {\Gamma'}/2}{ \Delta_{\rm A}' + \ii {\Gamma'}/2 + \sum_{j=1}^N ( J_{\rm 1D} + \ii \Gamma_{\rm 1D}/2  )\cos^2(x_j \pi/a)  }  \right|^2   . 
\end{align}
We have defined $\Delta_{\rm A}'\equiv\tilde{\Delta}_{\rm A}+\Delta_0$ in order to account for the AC-Stark shift $\Delta_0$ of the atoms due to the dipole trap.  

In order to accurately model the experimental conditions, we average the transmission model over atom positions and atom number.
During a single measurement, the atoms are free to move along the length of the device over the range $2\Delta x_A$ as in Fig. \ref{fig:trap}(a), evenly sampling the Bloch function.    We let $\langle T( \Delta_{\rm A} , N; x_1,...,x_N)  \rangle_x$ be an average over all positions, i.e.,
\begin{align*}  
\langle T( \Delta_{\rm A} , N; x_1,...,x_N)  \rangle_x = T_0(\Delta_{\rm A}) \int_0^{a} \mathrm{d}x_1 ...\mathrm{d}x_N \left|  \frac{\Delta_{\rm A}' + \ii {\Gamma'}/2}{ \Delta_{\rm A}' + \ii {\Gamma'}/2 + \sum_{j=1}^N ( J_{\rm 1D} + \ii \Gamma_{\rm 1D}/2  )\cos^2(x_j \pi/a)  }  \right|^2   . 
\end{align*}

We repeat the measurement multiple times for each frequency $\Delta_{\rm A}$.   Each experiment can have a different number of atoms, and so we average the transmission expression over a  Poisson distribution  $P_{\bar{N}}(N) $, which is a function of the average atom number $\bar{N}$.  The transmission model averaged over both atom positions and atom numbers is given by 
\begin{align} \label{eq:avgT}
&\langle T( \Delta_{\rm A} , N; x_1,...,x_N)  \rangle_{x,N} =\\\nonumber
& T_0(\Delta_{\rm A}) \sum_{N} \, P_{\bar{N}}(N)  \,\,  \langle T( \Delta_{\rm A} , N; x_1,...,x_N)  \rangle_x    . 
\end{align}
This is the final form of the transmission model that we use to fit the atomic spectra.   

\begin{figure*}[ht]
\centering
\includegraphics[width=0.8\linewidth]{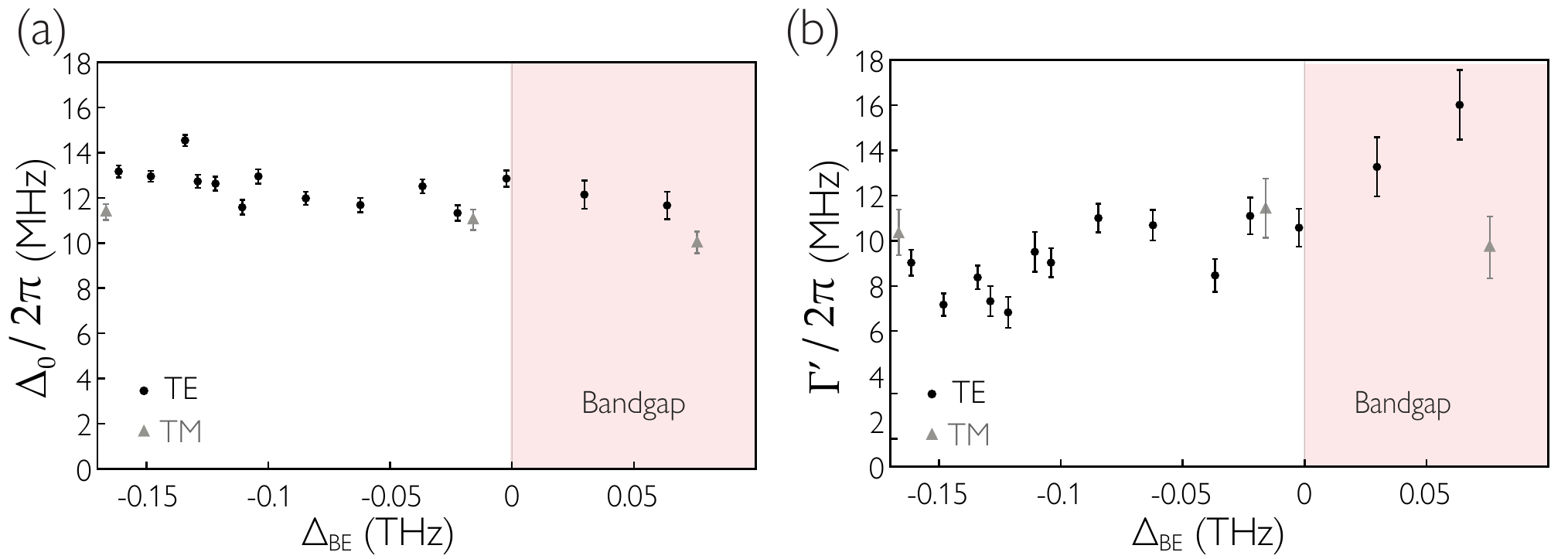} 
\caption[]{ Fitted values from averaged transmission model for TE (black, circles) and TM (gray, triangles) spectra. {\bf (a)}  Fitted AC Stark shift $\Delta_0$.  {\bf (b)}  Fitted $\Gamma'$.   } 
\label{fig:Delta0Gammap}
\end{figure*}

Assuming $\bar{N} = 3.0$, which is obtained from the atom decay rate measurement, we fit the TE atomic spectra with \eqref{eq:avgT} and extract  $\Gamma_{\rm 1D}$, $J_{\rm 1D}$, ${\Gamma'}$, and $\Delta_0$ for each frequency.  We show the values of $\Gamma_{\rm 1D}$ and $J_{\rm 1D}$ in Fig.~4(a) of the main text.  We show the AC Stark shift and non-guided decay rate in  Fig.~\ref{fig:Delta0Gammap}.    

The average of the non-guided decay rate $\Gamma'$ for the TE data outside the bandgap is $ {\Gamma}'= 2 \pi \times 9.1$~MHz. This is significantly larger than the expected value from the FDTD simulation, $\Gamma' =2 \pi \times  5.0$~MHz. This additional inhomogeneous broadening could be due to finite temperature of the trapped atoms, vector shifts from  circular light in the SI beam, atom density dependent collisional broadening, stray magnetic fields, and electric fields from charges in the dielectric.  We estimate the contributions individually, and find that they likely do not explain the extraneous broadening. We note that the estimate of `temperature' of trapped atoms could be improved in the future \cite{STLB08}, and it may help shed light on our excess broadening.  

Interestingly, the fitted ${\Gamma'}$ increases in the bandgap, and is as high as $\Gamma'= 2 \pi \times 16$~MHz for the last measured point. One possible explanation is that this is due to the break-down of the single bright mode approximation, as coupling to multiple collective atomic modes should result in a broadened linewidth. Another possibility is since there is a large extinction of the TE mode in the bandgap, there might be some mixing between the TE and TM modes. 

We also measure transmission spectra for the TM mode, whose band edges are far-detuned from the Cs transitions.  The transmission in this waveguide regime is described by an optical density model 
\begin{equation}
T/T_0 = \exp{\left[   \frac{- \text{OD}  }{1 + \left( \frac{2 \Delta_{\rm A}'}{ \Gamma_{\rm 1D}^{\rm TM} + {\Gamma'} } \right)^2}  \right]},
\label{eq:OD}
\end{equation}
where the resonant optical density is given by $ \text{OD} = 2\bar{N} \Gamma_{\rm 1D}^{\rm TM} / \tilde{\Gamma}' $.   We fit the TM spectra with this model and extract ${\Gamma}'$, $\Delta_0$, and $\Gamma_{\rm 1D}^{\rm TM}$ (assuming $\bar{N}=3$).     The values of ${\Gamma'}$ and $\Delta_0$ are shown with the TE data in Fig.~\ref{fig:Delta0Gammap}. The averaged $\Gamma_{\rm 1D}^{\rm TM}$ value is 0.044 $\Gamma_0$, which is $\sim 30$ times smaller than $\Gamma_{\rm 1D}$ for the TE mode at the first resonance $\nu_1$, and clearly demonstrates the enhanced interaction due to the PCW. 

\section{Simple transmission model}
\begin{figure*}[ht]
\centering
\includegraphics[width=0.8\linewidth]{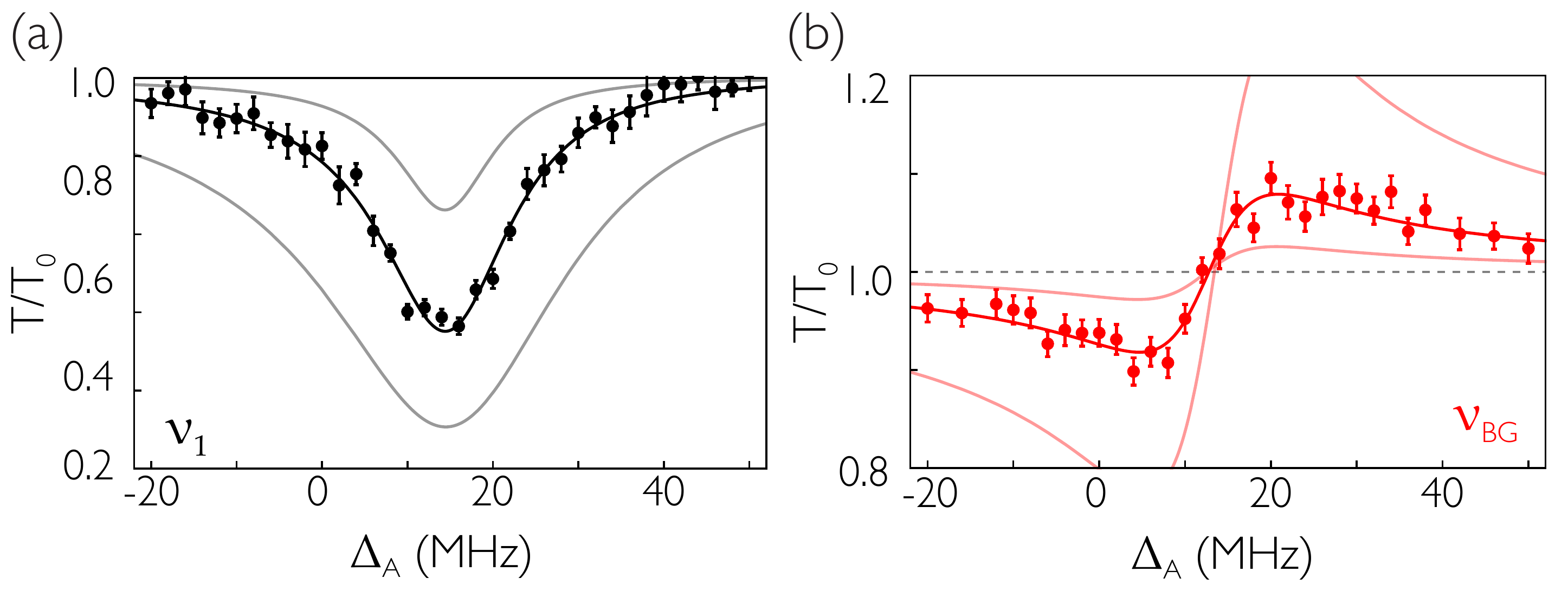} 
\caption[]{Fits of transmission spectra with model of \eqref{eq:TAB} for when the atomic resonance frequency is aligned to the first resonance {\bf (a)} and in the bandgap {\bf (b)}. From the decay rate measurement, the average number of atoms is $\bar{N}\approx 3 $, and the translucent curves give the expected spectra for $\bar{N}=1$ and $\bar{N}=9$ atoms.}    
\label{fig:ABspectra}
\end{figure*}
In the main text, we fit atomic transmission spectra with the averaged transmission model from \eqref{eq:avgT} in order to extract the peak guided-mode decay rate $\Gamma_{\rm 1D}$ and frequency shift $J_{\rm 1D}$.  
In this section, we fit the spectra with a transmission model which involves no averaging, and we extract an effective decay rate $\Gamma_{\rm 1D}^{\rm eff}$ and frequency shift $J_{\rm 1D}^{\rm eff}$, which will be smaller than the corresponding peak values due to the averaging of the $\cos^2(\pi x/a)$ Bloch function as the atoms move along the $x$-axis of the trap.   In the ``single-bright-mode" approximation discussed in the main text, the transmission for a single collective mode with total decay rate $A$ and frequency shift $B$ is given by 
\begin{equation}  \label{eq:TAB}
\frac{  T( \Delta_{\rm A}) }{ T_0(\Delta_{\rm A})}= \left|  \frac{\Delta_{\rm A}' + \ii {\Gamma'}/2}{ \Delta_{\rm A}'+B + \ii (\Gamma'+A)/2 }  \right|^2   . 
\end{equation}
Here, the detuning $\Delta_A'$ includes the AC stark shift $\Delta_A' = \Delta_A + \Delta_0 $.  Since the average number of atoms $\bar{N} \approx 3$ is measured independently in a decay rate measurement, the collective rates $A$ and $B$ are related to the effective rates by  $A = \bar{N} \Gamma_{\rm 1D}^{\rm eff}$ and $B = \bar{N} J_{\rm 1D}^{\rm eff}$. Examples of the fitted spectra for atoms outside and inside the band-gap are shown in Fig.~\ref{fig:ABspectra}.  The translucent lines are the expected signals for an average atom  number of $\bar{N}=1$ and  $\bar{N}=9$.  

\begin{figure*}[ht]
\centering
\includegraphics[width=0.8\linewidth]{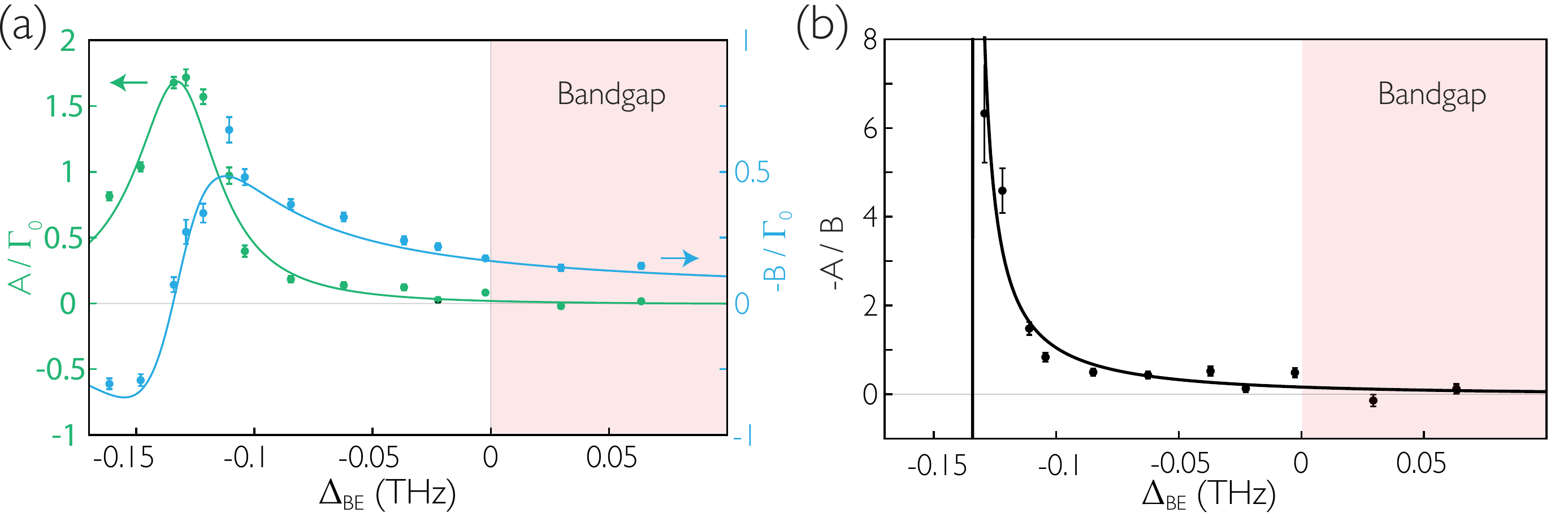} 
\caption[]{{\bf (a)} Fitted values for the effective collective decay rates $A$ and frequency shifts $B$ for various detunings from the band-edge $\Delta_{\rm BE}$.  The solid lines are the expected result for the peak values, except scaled down by $\eta = 0.42$.  {\bf (b)}  Ratio $A/B = \Gamma_{\rm 1D}^{\rm eff}/J_{\rm 1D}^{\rm eff}$, along with the theoretical prediction for the peak ratio $\Gamma_{\rm 1D}/J_{\rm 1D}$ from Fig.~4(b) of the main text.} 
\label{fig:ABModel}
\end{figure*}

The fitted values of $A$ and $B$ are plotted for each detuning from the band-edge $\Delta_{\rm BE}$ in Fig.~\ref{fig:ABModel}(a). The results are qualitatively similar to the corresponding plot in Fig.~4(a) in the manuscript, except the effective rates $A= \bar{N} \Gamma_{\rm 1D}^{\rm eff}$ and $B= \bar{N} \Gamma_{\rm 1D}^{\rm eff}$ are scaled down by $\eta = 0.42 $ due to the modulation of the Bloch function  $\cos^2(\pi x/a) $.   The solid line in Fig.~\ref{fig:ABModel}(a) is the same theoretical curve as in Fig~4(a) except scaled by $\eta =0.42$. 

The ratio of $A/B = \Gamma_{\rm 1D}^{\rm eff}/J_{\rm 1D}^{\rm eff}$ is plotted in Fig.~\ref{fig:ABModel}(b).  Since the scale factors $\eta$ cancel, the result is in good agreement with the corresponding plot of $\mathcal{R}=\Gamma_{\rm 1D} / J_{\rm 1D}$ in Fig.~4(b) of the manuscript.   The black theory curve is the same as in the manuscript.  Whereas the peak decay rate and frequency shift is sensitive to the specific model, the ratio of dissipative to coherent coupling is mostly model insensitive. 

\section{Atom decay measurement}
\begin{figure*}[ht]
\centering
\includegraphics[width=0.5\linewidth]{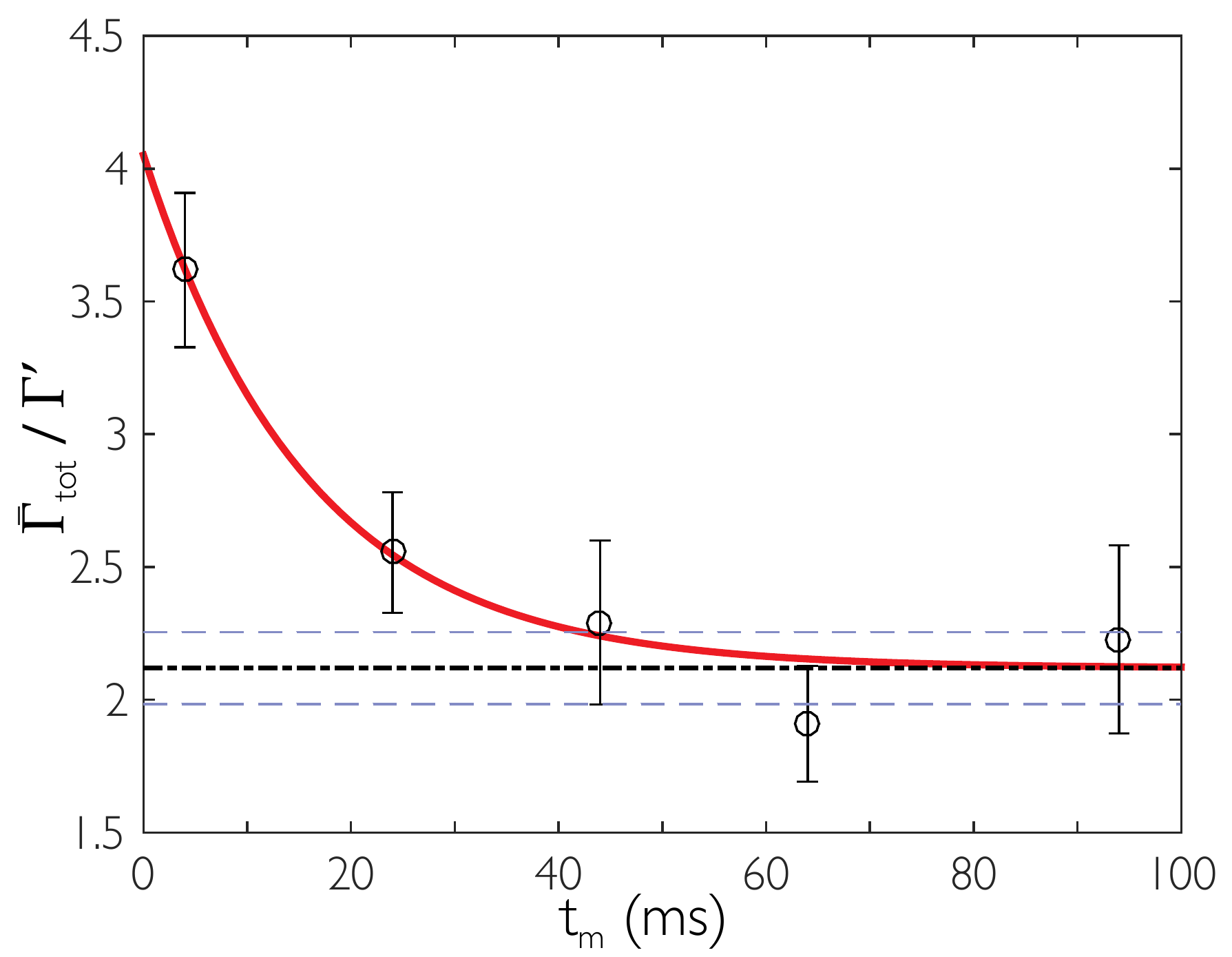} 
\caption{Total decay rates as a function of holding time $t_{\rm  m}$. The red solid curve is the empirical fit and the dash-dot line represents the fitted asymptotic total decay rate at very long times. The blue dashed lines specify fitted error boundaries. The fit yields $\tau_{\rm  SR}  = 16$ ms, $\bar{\Gamma}_{\rm  SR} = 1.5\Gamma^{'}$ and the asymptote $\bar{\Gamma}^{(1)}_{\rm  tot}/\Gamma^{'} = 2.12\pm0.14$.} 
\label{fig:atomdecay2}
\end{figure*}
We exploit the superradiance of atoms trapped near the alligator PCW to determine the mean atom number $\bar{N}$ and the peak atom decay rate $\Gamma_{\rm  1D}$ (at $\nu_1$) into the guided-modes.  

As established in Ref. \cite{SGHH15}, the total exponential decay rates of the atoms is $\bar{\Gamma}_{\rm  tot}(\bar{N}) = \bar{\Gamma}_{\rm  SR}(\bar{N})+\bar{\Gamma}^{(1)}_{\rm  tot}$, where $\bar{\Gamma}_{\rm  SR}$ is the $\bar{N}$-dependent superradiance decay rate, and $\bar{\Gamma}^{(1)}_{\rm  tot}$ is the observed single-atom decay rate.  We note that when $\bar{N}\ll 1$, $\bar{\Gamma}_{\rm  tot }\sim \bar{\Gamma}^{(1)}_{\rm  tot} = \bar{\Gamma}_{\rm  1D} +\Gamma^{'}$,  since
only the single-atom decay rate into GM $\bar{\Gamma}_{\rm  1D} $ and into environment $\Gamma^{'}$ remain. $\Gamma^{'}$ is numerically calculated to be $2\pi \times 5.0$ MHz for cesium D$_1$ line at the trapping site near the PCW \cite{SGHH15}.

We excite the atoms with a weak resonant light pulse through the guided-mode, while the first resonance $\nu_1$ near the band edge is aligned with cesium D$_1$ line. Pulse properties are as in Ref. \cite{SGHH15}.  The subsequent fluorescence decay rates $\bar{\Gamma}_{\rm  tot}$ are determined through exponential fits.  By varying the trap holding time $t_{\rm  m}$ after loading, the mean atom numbers for the decay measurements are varied.  The decay rates are empirically fitted in an exponential form as a function of holding time $t_{\rm  m}$~\cite{SGHH15}: $\bar{\Gamma}_{\rm  tot}(t_{\rm  m})=\bar{\Gamma}_{\rm  SR}e^{-t_{\rm  m}/\tau_{\rm  SR}} + \bar{\Gamma}^{(1)}_{\rm  tot}$, as shown in Fig.~\ref{fig:atomdecay2}. 
From the fitted asymptotic-value of the decay rates, we deduce that the apparent single-atom decay rate is  $\bar{\Gamma}_{\rm  1D} = (1.12\pm0.14 )\Gamma^{'}$.

Because the atoms are randomly distributed along $x$ direction in the trap,  the observed decay curves are results after spatial averaging the coupling rates $\Gamma_{\rm  1D}(x)$.  Assuming an uniform distribution of $N$ atoms around the center of the PCW, a more detailed model specifies the form of fluorescence intensity decay as \cite{SGHH15}:  
\begin{equation}  \label{eqn:I}
\begin{split}
\mathcal{I}_N(t) =\gamma^2 e^{-(N\gamma+\Gamma')t}\cdot{I_{0}}\left(\gamma t\right)^{N-2}\cdot   \Big[ \frac{N(N+1)}{4}{I_0}\left(\gamma t\right)^2\\
-\left(\frac{N}{4 \gamma t} +\frac{N^2}{2}\right){I_{0}}\left(\gamma t\right){I_{1}}\left(\gamma t\right) 
+\frac{N(N-1)}{4}{I_{1}}\left(\gamma t\right)^2 \Big],
\end{split}
\end{equation}
where $\gamma = \Gamma_{\rm  1D}/2$, and $I_{k}$ is the modified Bessel function. Numerically simulating the decay of single atoms in the trap by using $\mathcal{I}_1(t)$, we compare between the exponentially fitted value $\bar{\Gamma}_{\rm  1D}$ and  the value of ${\Gamma}_{\rm  1D}$ used for $\mathcal{I}_1(t)$, which yields a ratio of $\bar{\Gamma}_{\rm  1D}/\Gamma_{\rm  1D} = 0.81$. This is consistent with the ratio of $0.8\pm0.3$ from measurement at long hold time $t_{\rm  m}= 94$ ms, when single-atom decay predominates (shown as the asymptote in Fig.~\ref{fig:atomdecay2}). Based on the values of $\bar{\Gamma}_{\rm  1D}$ deduced above, we conclude that $\Gamma_{\rm  1D} = (1.4\pm 0.2) \Gamma^{'}$.

At early holding times, the atom number $N$ noticeably fluctuates around some mean values $\bar{N} \gtrsim 1$ .  To capture this $\bar{N}$-dependent variation, we fit the decay curves by averaging $\mathcal{I}_N(t)$ with weight function of Poisson distribution  probability $P_{\bar{N}}(N)$ \cite{SGHH15}.  The fitting parameter here is $\bar{N}$, while we fix the value of $\Gamma_{\rm  1D}$ in Eq.~\ref{eqn:I}.  The fit is consistent with $\bar{N} = 3.0 \pm 0.5$ at $t_{\rm  m} = 4$ ms when we carry out the transmission spectra measurement. Based on the trap life time $\tau = 30$ ms, we further deduce that  $\bar{N} \sim 0.1$ at $t_{\rm  m}=94$ ms . 

The linear $\bar{N}$-dependence of superradiance is given by $\bar{\Gamma}_{\rm  SR}=\eta\cdot\bar{N}\cdot\Gamma_{\rm  1D}$, where $\eta = 0.36\pm0.06$ is some linear coefficient, whose value is consistent with that reported in Ref. \cite{SGHH15}.

\end{document}